\documentclass[journal]{IEEEtran}

\usepackage{array,graphicx,cite,epsfig,amssymb,amsfonts}
\usepackage{color,bbm,amsmath,epstopdf,algorithm,algorithmic,subfigure, multirow,url,mathtools}
\usepackage{tabularx,comment,enumerate, caption}

\newcommand{\be}{\begin{equation}}
\newcommand{\ee}{\end{equation}}
\def\bee#1\eee{\begin{align}#1\end{align}}
\newcommand{\bse}{\begin{subequations}}
\newcommand{\ese}{\end{subequations}}
\newcommand{\nnb}{\nonumber}

\newtheorem{theorem}{\textbf{Theorem}}

\newtheorem{fact}{\textbf{Fact}}
\newtheorem{corollary}{\textbf{Corollary}}

\newtheorem{definition}{\textbf{Definition}}

\newcommand{\red}[1]{{\color{red}#1}}

\def \ISTR {}

\begin{document}

\title{Device-to-Device Load Balancing for Cellular Networks}

\author{Lei Deng,
        Yinghui He,
        Ying Zhang,
        Minghua~Chen, \\
        Zongpeng Li,
        Jack Y. B. Lee,
        Ying Jun (Angela) Zhang,
        and~Lingyang Song

\thanks{
The work presented in this paper was supported in part by the University Grants Committee of the Hong Kong Special Administrative Region, China (Collaborative Research Fund No. C7036-15G), in part by NSFC (Project No. 61571335 and 61628209), and in part by Hubei Science Foundation (Project No. 2016CFA030 and 2017AAA125).
Part of this work has been presented at IEEE MASS, 2015 \cite{deng2015device}.
\emph{(Corresponding author: Minghua Chen.)}}
\thanks{L. Deng is with the School of Electrical Engineering \& Intelligentization, Dongguan University of Technology, Dongguan 523808, China (email: denglei@dgut.edu.cn).}
\thanks{Y. He is with the College of Information Science and Electronic Engineering, Zhejiang University, Hangzhou 310027, China (e-mail: 2014hyh@zju.edu.cn).}
\thanks{Y. Zhang, M. Chen, J. Lee, Y. Zhang are with the Department of Information Engineering, the Chinese University of Hong Kong, Hong Kong, China (e-mail:
ying.ie.cuhk@gmail.com; minghua@ie.cuhk.edu.hk; jacklee@computer.org; yjzhang@ie.cuhk.edu.hk).}
\thanks{Z. Li is with School of Computer Science, Wuhan University, 299 Baiyi Road, Wuhan, Hubei 430072, China (e-mail: zongpeng@whu.edu.cn).}
\thanks{L. Song is with the School of Electrical Engineering and Computer Science, Peking University, Beijing 100871, China
(e-mail:  lingyang.song@pku.edu.cn).}
}


\maketitle

\begin{abstract}

Small-cell architecture is widely adopted by cellular network operators to increase spectral spatial efficiency.
However, this approach suffers from low spectrum temporal efficiency. When a cell becomes smaller and covers fewer users, its total traffic fluctuates significantly due
to insufficient traffic aggregation and exhibits a large ``peak-to-mean" ratio. As operators customarily provision spectrum for
peak traffic, large traffic temporal fluctuation inevitably leads to low spectrum temporal efficiency.
To address this issue, in this paper, we advocate device-to-device (D2D) load-balancing as a useful mechanism.
The idea is to shift traffic from a congested cell to its adjacent under-utilized cells by leveraging inter-cell D2D communication, so that the traffic can be served
without using extra spectrum, effectively improving the spectrum temporal efficiency.
We provide theoretical modeling and analysis to characterize the benefit of D2D load balancing, in terms of total
spectrum requirements and the corresponding cost, in terms of incurred D2D traffic overhead.
We carry out empirical evaluations based on real-world 4G data traces and show that D2D load balancing can reduce the spectrum requirement by 25\% as compared to the
standard scenario without D2D load balancing, at the expense of negligible 0.7\% D2D traffic overhead.
\end{abstract}


\IEEEpeerreviewmaketitle

\section{Introduction} \label{sec:introduction}

\IEEEPARstart{T}{he} drastic growth in mobile devices and applications has triggered
an explosion in cellular data traffic. According to Cisco~\cite{cisco17},
global cellular data traffic
reached $7$ exabytes per month in 2016
 and will further witness a 7-fold increase
in 2016-2021. Meanwhile, radio frequency remains a scarce resource
for cellular communication.
Supporting the fast-growing data traffic demands has become a central
concern of cellular network operators.

There are mainly two lines of efforts to address this concern. The
first is to serve cellular traffic by exploring additional
spectrum, including offloading cellular traffic to WiFi \cite{wifi10}
and the recent 60GHz millimeter-wave communication endeavor \cite{60G10}.
The second is to improve \emph{spectrum spatial efficiency}. A common approach
is to adopt a small-cell architecture, such as micro/pico-cell\cite{smallcell12}. By reducing cell size, operators can pack
more (low-power) base stations in an area and reuse radio frequencies
more efficiently to increase network capacity.

While the small-cell architecture improves the \emph{spectrum spatial
efficiency}, it comes at a price of degrading
the \emph{spectrum temporal efficiency}. When a cell becomes smaller
and covers fewer users, there is less traffic aggregation. Consequently,
the total traffic of a cell fluctuates significantly,
exhibiting a large ``peak-to-mean\textquotedblright{}
ratio. As operators customarily provision spectrum to a
cell based on peak traffic, high temporal fluctuation in traffic volumes
inevitably leads to low spectrum temporal efficiency.

To see this concretely, we carry out a case-study based on 4G
cell-traffic traces from Smartone \cite{smartone} (this complements the study in our conference version~\cite{deng2015device}, which was based on 3G data traces),
a major cellular network operator
in Hong Kong, a highly-populated metropolis.
The detailed analysis and description can be found
\ifx \ISTR \undefined
in our technical report \cite{TR}.
\else
in Appendix \ref{app:case-study}.
\fi
Based on this case study, we observe that the average cell-capacity
utilization is very now and the peak traffic of many pairs of adjacent BSs occurs at different time epochs.
This confirms that small-cell architecture indeed causes very low
spectrum temporal utilization, and it suggests ample room to do traffic load balancing to improve temporal utilization.

Motivated by the above observations, we advocate \emph{device-to-device
(D2D) load-balancing} as a useful mechanism to improve spectrum temporal
efficiency. D2D communication~\cite{Doppler09} \cite{Foder12} is a promising paradigm for
improving system performance in next generation cellular networks
that enables direct communication between user devices  using cellular frequency. It is conceivable to
relay traffic from congested cells to adjacent
underutilized cells via inter-cell D2D communication, enabling load-balancing
across cells at the expense of incurred inter-cell D2D traffic.

We remark that an idea of this kind was also studied by Liu \emph{et al.} in their recent work~\cite{Liu14}. They focus on important
aspects of examining the technical feasibility of D2D load balancing and practical algorithm design
in three-tier LTE-Advanced networks. This work is complement to
their study and focuses on the following two important questions:
\begin{itemize}
\item How much spectrum reduction can D2D load balancing bring to a cellular network?
\item What is the corresponding D2D traffic overhead for achieving the benefit?
\end{itemize}
Answers to these questions provide fundamental understanding of the
viability of D2D load balancing in cellular networks. In this paper,
we answer the questions via both theoretical analysis and
empirical evaluations based on real-world traces. We make the following
contributions.

$\mbox{\ensuremath{\rhd}}$ In Sec.~\ref{sec:example}, using perhaps
the simplest possible example, we illustrate the concept of D2D load
balancing and show that it can reduce peak traffic for two adjacent
cells by 33\%. We also compute the associated D2D traffic overhead.

$\mbox{\ensuremath{\rhd}}$ For general settings beyond the example,
we provide tractable models to analyze
the performance of D2D load balancing in Sec. \ref{sec:system_model}.
We also exploit the optimal solutions without and with D2D
load balancing in Sec. \ref{sec:optimal_no_d2d} and Sec. \ref{sec:optimal_d2d},
respectively.

$\mbox{\ensuremath{\rhd}}$
Theoretically, for arbitrary settings,
we derive an upper bound for the benefit of D2D
load balancing, in terms of sum peak traffic reduction
in
Sec.~\ref{sec:a_general_upper_bound}.
We show that the bound is asymptotically tight for a specified network scenario,
where we further derive the corresponding overhead, in terms of incurred D2D
traffic.
Our bound and analysis reveal the insight behind the effectiveness of D2D
load balancing: by aggregating traffic among
adjacent cells via inter-cell D2D communication, we can leverage
statistical multiplexing gains to better serve the overall traffic
without requiring extra network capacity.

$\mbox{\ensuremath{\rhd}}$
Empirically, in Sec.~\ref{sec:simulation}, we use real-world 4G data traces to
verify our theoretical analysis and reveal that D2D load balancing
can reduce sum peak traffic of individual cells by 25\%, at the cost
of 0.7\% D2D traffic overhead. This implies significant spectrum saving at a negligible system overhead.


Throughout this paper, we assume that time is slotted into intervals of unit length, and
each wireless hop incurs one-slot delay. We focus on uplink communication scenarios,
while our analysis is also applicable to
the downlink communication. In addition,
in the rest of this paper, for any two positive integers $K_1, K_2$ with $K_1 < K_2$, we use notation $[K_1,K_2]$ to denote set $\{K_1,K_1+1,\cdots,K_2\}$,
i.e., $[K_1,K_2] \triangleq \{K_1,K_1+1,\cdots,K_2\}$. When $K_1=1$, we further simplify notation $[1,K_2]$ to be $[K_2]$, i.e., $[K_2] \triangleq \{1,2,\cdots, K_2\}$.
\ifx \ISTR \undefined
Due to the page limit, we move some of our technical proofs and some discussions into our technical report \cite{TR}.
\else
\fi

\section{Related Work}

In this paper, we use a dataset from Smartone to show that the peak traffic of different adjacent BSs
occurs at different time epochs.
Similar observation is also obtained from the measurement studies in \cite{Hu2015Indepth} and \cite{xu2017understanding}.
The authors in \cite{Hu2015Indepth} analyze the 3G cellular traffic of three major cities in China during 2010 and 2013
and a city in a Southeast Asian country in 2013. They show that the correlation coefficient of the traffic profiles of different BSs
is small (between 0.16 and 0.33).
The authors in \cite{xu2017understanding} analyze the 3G/4G cellular traffic of 9600 BSs in Shanghai, China in 2014.
They show that different areas (residential area, business district, transport, entertainment, and
comprehensive area) have different traffic patterns, which have different peak epochs.
All these traffic measurements motivate us to do load balancing among different BSs
so as to reduce the peak demand (spectrum requirement).

In this paper, we propose the D2D load balancing scheme to reduce the peak demand (spectrum requirement)
of BSs. There are other load balancing schemes to achieve the goal, including
smart user association \cite{andrews2014overview,ye2013user} and mobile offloading \cite{lee2014economics}.


Smart user association \cite{andrews2014overview,ye2013user} dynamically associates users to the BSs so as to balance the traffic demand of all BSs.
However, (i) smart user association schemes normally should be operated on large timescale to overcome the large overhead incurred
by frequently switching from one BS to another BS (\emph{a.k.a.}, handover) \cite{andrews2014overview}; thus it is not designed
for balancing traffic across BSs on small timescale, and (ii) smart user association scheme in~\cite{ye2013user}, where cellular operators globally
associate every user to a BS in a centralized manner, incurs high overhead and complexity. Other smart user association schemes through cell breathing~\cite{Jalali98breathing} or power control methods, where every user locally connects to the BS with strongest signal in a distributed manner,
will change the interference levels significantly and thus they may need for spectral re-allocation across the whole networks.
Instead, D2D scheme can do load balancing on short timescale since D2D communications often occur locally within short distances and low power
and thus D2D scheme has limited impact to the cellular network.
Although D2D load balancing may need to switch between the BS mode (connecting to the BS) and the D2D mode (connecting to the device),
such a switch happens locally and it is more lightweight than the global handover between different BSs.
Therefore, though D2D load balancing scheme will incur some overhead during D2D communications, it has some unique
advantages over smart user association schemes.
Meanwhile, we also remark that
D2D scheme and smart user association schemes are  complementary
for load balancing in the sense that we might simultaneously use smart user association schemes on large timescale and
use D2D scheme on small timescale. Thus, in this paper we advocate the D2D load balancing scheme.

Mobile offloading \cite{lee2014economics,dimatteo2011cellular,han2011mobile,wifi10} is another scheme to reduce the cellular traffic demand. It mainly uses  WiFi infrastructure.
However, mobile offloading and D2D load balancing are technically
different schemes: mobile offloading aims to exploit
outband spectrum, but our D2D load balancing scheme targets to increase inband cellular
temporal spectrum efficiency. Furthermore in D2D load balancing,
the cellular operation can ubiquitously control everything, including both
D2D and user-to-BS transmissions. However, mobile offloading
usually outsources a portion of traffic to a thirdparty
entity, imposing unpleasant unreliability for transmissions.
Therefore, our proposed D2D load balancing scheme can ensure better QoS than mobile offloading.
Again, our D2D load balancing scheme are orthogonal to the mobile offloading scheme
in the sense that the operators can simultaneously use them to reduce the cellular spectrum requirement.

In addition to those traffic load balancing schemes, spectrum reallocation is another effective approach to reduce the spectrum requirement.
Instead of moving traffic among different cells, spectrum reallocation
\emph{dynamically} allocate the spectrum among different cells to better match
the time-varying traffic demands
\cite{zhuang2017scalable,zhuang2015traffic,zhou2017licensed,Zhou20171000Cell}.
However, spectrum allocation incurs high complexity. The state-of-the-art spectrum allocation
solution is proposed in \cite{Zhou20171000Cell}, which can
obtain near-optimal performance for a network with up  to 1000 APs and 2500 active users.
Furthermore, spectrum reallocation again is operated on large timescale. Hence,
the cellular operator can simultaneously do spectrum reallocation
on large timescale based on aggregated traffic information \cite{zhuang2015traffic}
and use our proposed D2D load balancing scheme on small timescale based on the fine-grained traffic information
to reduce the spectrum requirement.

We further remark that there are some existing works on D2D load balancing.
For the three-tier LTE-Advanced heterogenous networks,
\cite{Liu14} examines the technical feasibility and designs practical algorithm for D2D load balancing;
\cite{chen2015load,vlachos2015optimal,jiang2017relay} propose research allocation strategies to
achieve load balancing goal via D2D transmission.
In \cite{hajiesmaili2017incentivizing}, an auction-based mechanism is proposed to incentivize
the mobile users to participate in D2D load balancing.
However, all existing works do not directly answer the two important questions proposed in Sec.~\ref{sec:introduction}.

\section{An Illustrating Example} \label{sec:example}

\begin{figure*}
\centering \subfigure[Cellular network topology and traffic demands.]{
\label{fig:example_topology}
\includegraphics[width=0.9\linewidth]{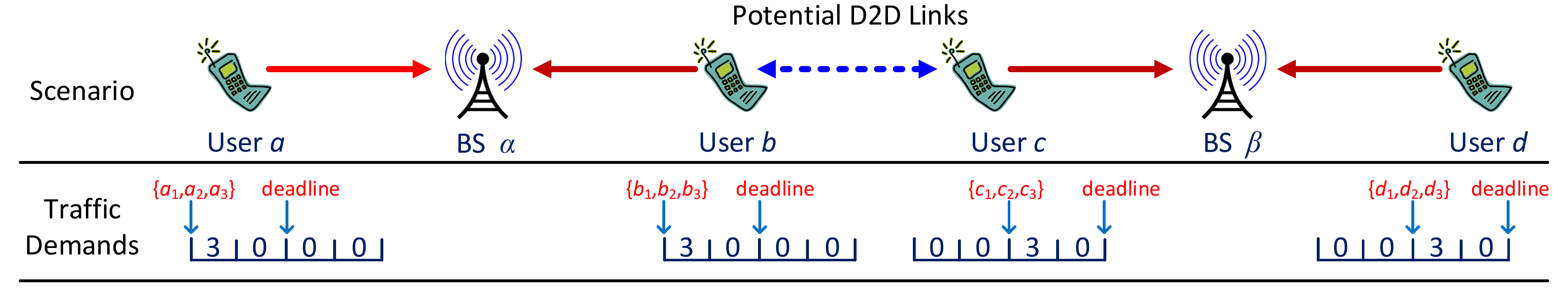}}
\\
 \subfigure[Conventional cellular approach without D2D.]{ \label{fig:example_no_d2d}
\includegraphics[width=0.9\linewidth]{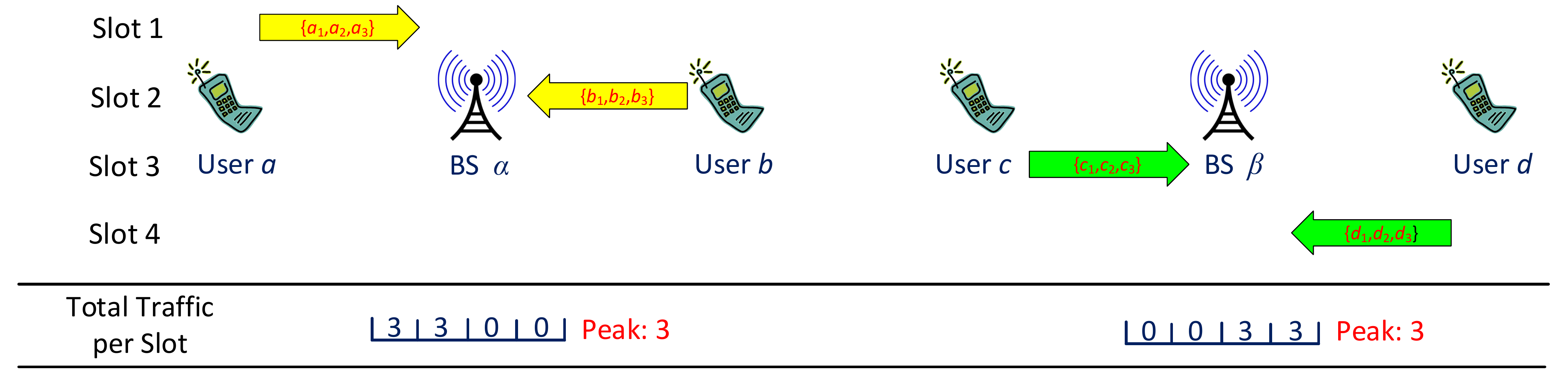}}\\
 \subfigure[Our approach with D2D load balancing.]{ \label{fig:example_d2d}
\includegraphics[width=0.9\linewidth]{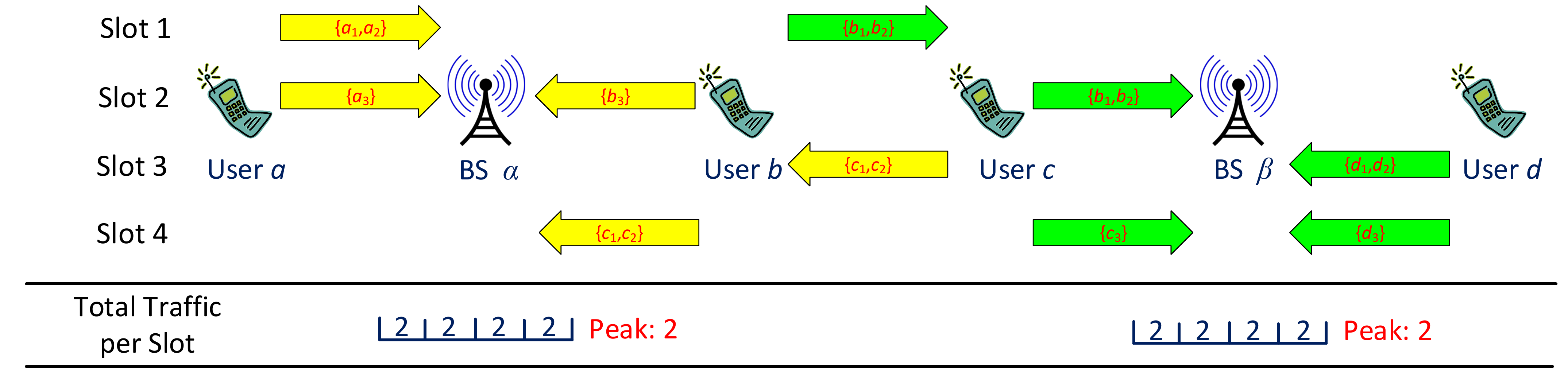}}
\protect\caption{A simple example for demonstrating the concept of D2D load balancing,
and that it can reduce the peak traffic for both cells by 33\% (both
from 3 to 2) at the cost of 4 extra inter-cell D2D transmissions.}
\label{fig:example} 
\end{figure*}

We consider a simple scenario shown in Fig. \ref{fig:example_topology},
where 4 users are each aiming at transmitting 3 packets to two base stations
(BS) subject to a deadline constraint. We compare the peak traffic of
both BSs for the case without D2D load balancing (Fig. \ref{fig:example_no_d2d})
and for the case with D2D load balancing (Fig. \ref{fig:example_d2d}).
We illustrate the concept of D2D load balancing and show that it can
reduce the peak traffic for two adjacent cells by 33\%.

Specifically, we consider a cellular network of two adjacent cells served by BS $\alpha$ and BS $\beta$, and four users
$a$, $b$, $c$, $d$. BS $\alpha$ (resp. $\beta$)
can directly communicate with only users $a$ and $b$ (resp. users
$c$ and $d$). BS $\alpha$ and BS $\beta$ use orthogonal frequency bands. Due to proximity,
users $b$ and $c$ can communicate with each other using frequency band
of either BS $\alpha$ or $\beta$, creating inter-cell D2D links.
Both user $a$ and user $b$  generate 3 packets at the beginning of slot 1, and
both user $c$ and user $d$ generate 3 packets
at the beginning of slot 3. All packets have the same size and a delay constraint of
2 slots, i.e., a packet must reach BS $\alpha$ or $\beta$
within 2 slots from its generation time. {\em We assume that
a packet is successfully delivered as long as it reaches any BS},
since BSs today are connected by a high-speed optical backbone, supported by
power clusters, and can coordinate to jointly
process/forward packets for users.

In the conventional approach without D2D load balancing, a user only
communicates with its own BS. It is straightforward to verify that
the minimum peak traffic of both BS $\alpha$ and BS $\beta$
is 3 (unit: packets), and can be achieved by the scheme in Fig.
\ref{fig:example_no_d2d}. For instance, the minimum peak traffic
for BS $\alpha$ is achieved by user $a$ (resp. user $b$) transmitting
all its 3 packets to BS $\alpha$ in slot 1 (resp. slot 2).

With D2D load balancing,
we can exploit the inter-cell D2D links between users $b$ and $c$ to perform load balancing
and reduce the peak traffic for both BS $\alpha$ and BS $\beta$.
\begin{itemize}
\item
In slot 1, user $a$ transmits two packets $a_{1}$ and $a_{2}$ to
BS $\alpha$, and user $b$ transmits two packets $b_{1}$ and $b_{2}$
to user $c$ using the orthogonal frequency band of BS $\beta$. The
traffic is 2 for both cells. In slot 2, users $a$ and $b$ transmit
their remaining packets $a_{3}$ and $b_{3}$ to BS $\alpha$, and
user $c$ relays the two packets it received in slot 1, i.e.,
$b_{1}$ and $b_{2}$, to BS $\beta$. The traffic is again 2 for
both cells. By the end of slot 2, we deliver
6 packets for users $a$ and $b$ to BSs.
\item In slots 3 and 4, note that users $c$ and $d$ have the same traffic
pattern as users $a$ and $b$, but offset by 2 slots. Thus we can
also deliver 3 packets for both users $c$ and $d$ in two slots. The traffic
of both BSs is 2 per slot.
\end{itemize}
Overall, with D2D load balancing, we can serve all traffic demands with
peak traffic of 2 for both BSs, which is 33\% reduced as compared
to the case without D2D load balancing.

The intuition behind this example is that the peak traffic for the
two cells occurs at different time instances. When users $a$ and $b$
transmit data to BS $\alpha$ in the first two slots, BS $\beta$
is idle. Meanwhile, BS $\alpha$ is idle when users $c$ and
$d$ transmit data to BS $\beta$ in the last two slots. Therefore,
D2D communication can help load balance traffic from the busy BS
to the other idle BS, reducing the peak traffic for both BSs.
However, D2D load balancing also comes with cost, since it requires transmissions
over the inter-cell D2D links. In the example, the total traffic is
$8\times2=16$ packets and the D2D traffic is $2\times2=4$ packets, yielding
an overhead traffic ratio of $\frac{4}{16}=25\%$. Such D2D traffic
is the overhead that we pay in return for peak traffic reduction.

\section{System Model} \label{sec:system_model}
In this section, we present the system model for a general network topology
and a general traffic demand model beyond the simple example expounded
in the previous section. Such models will be used to analyze the benefit
of D2D load balancing in general settings, in terms of spectrum
reduction ratio, and the cost in terms of D2D traffic overhead ratio.

\subsection{Cellular Network Topology} \label{sec:topology_model}
Consider an uplink wireless cellular network with multiple cells and multiple mobile users.
We assume that each cell has one BS and each user is associated with one BS\footnote{We say that user $u$ is associated with BS $b$ if user $u$ is
in the cellular cell covered by BS $b$. When a user is covered by multiple BSs,
we assume that this user has been associated with one of them, e.g., the one with the strongest signal-to-noise ratio.
In the rest of this paper,
we will also use the terminology, cell $b$, to represent the cell covered by BS $b$.}.
Define $\mathcal{B}$ as the set of all BSs,
$\mathcal{U}_b$ as the set of users belonging to BS $b \in \mathcal{B}$,
and $\mathcal{U} = \cup_{b \in \mathcal{B}} \mathcal{U}_b$ as
the set of all users in the cellular network.
Let $b_u \in \mathcal{B}$ denote the cell (or BS) with which user $u \in \mathcal{U}$ is associated.
We model the uplink cellular network topology as a directed graph
$\mathcal{G}=(\mathcal{V},\mathcal{E})$ with vertex set
$\mathcal{V} = \mathcal{U} \cup \mathcal{B}$ and edge set
$\mathcal{E}$ where $(u,v) \in \mathcal{E}$ if there is
a wireless link from vertex (user) $u \in \mathcal{U}$ to vertex (BS or user) $v \in \mathcal{V}$.

\subsection{Traffic Model} \label{sec:traffic_model}
We consider a time-slotted system with $T$ slots in total, indexed from 1 to $T$.
Each user can generate a delay-constrained traffic demand at the beginning of any slot.
We denote $\mathcal{J}$ as the demand set. Each demand $j \in \mathcal{J}$ is characterized by the tuple $(u_j, s_j, e_j, r_j)$
where
\begin{itemize}
\item $u_j \in \mathcal{U}$ is the user that generates demand $j$;
\item $s_j \ge 1$ is the starting time/slot of demand $j$;
\item $e_j \in [s_j, T]$ is the ending time/slot (deadline) of demand $j$;
\item $r_j > 0$ is the volume of demand $j$ with unit of bits.
\end{itemize}
Namely, demand $j$ is generated by user $u_j$ at the beginning of slot $s_j$
with the volume of $r_j$ bits and it must be delivered to BSs before/on the end of slot $e_j$,
implying a \emph{delay} requirement $(e_j - s_j +1)$.
We also call interval $[s_j, e_j]$ the \emph{lifetime} of the demand $j$.
We further denote $\mathcal{J}_b$
as the set of demands that are generated by the users in BS $b \in \mathcal{B}$, i.e.,
$
\mathcal{J}_b \triangleq \{j \in \mathcal{J}: u_j \in \mathcal{U}_b\}.
$
Demand $j$ is delivered in time if every bit of demand $j$ reaches a BS
before/on the end of slot $e_j$. Note that different bits in demand $j$ could reach different BSs.
Thus, every user can transmit a bit either to its own BS directly in a single hop or to another user
via the D2D link between them such that the bit can reach another BS in multiple hops.

\subsection{Wireless Channel/Spectrum Model}
For each link $(u,v)\in\mathcal{E}$, we
denote its link rate as $R_{u,v}$ (units: bits per slot per Hz), which
is the number of bits that can be transmitted in one unit (slot) of time resource and with
one unit (Hz) of spectrum resource. Then if we allocate $x \in \mathbb{R}^+$ (unit: Hz) spectrum to link $(u,v)$ at slot $t$,
this link can transmit $x\cdot R_{u,v}$ bits of data from node $u$ to node $v$ in slot $t$.
Note that we simplify the channel model by assuming a linear relationship between the allocated spectrum
and the transmitted data. This assumption is reasonable for the high-SNR scenario when we use
Shannon capacity as the link rate \cite{shannon}.
In addition, we assume that the total spectrum is not divided into uplink spectrum and downlink
spectrum. Instead, our scheme allocates spectrum from a spectrum pool to mobile users for transmitting or
receiving data. Thus, in this paper, we do not consider the switching issue between uplink spectrum and downlink spectrum.

\subsection{Performance Metrics} \label{sec:performance_metrics}
In this paper, we aim at minimizing the total (amount of) spectrum to deliver all demands in $\mathcal{J}$ in time.
In particular, we need to obtain the minimum spectrum/frequency to serve all demands in time without D2D (resp. with D2D), denoted by $F^{\textsf{ND}}$
(resp. $F^{\textsf{D2D}}$).
To evaluate the impact of D2D load balancing, we characterize both the benefit and the cost for D2D load balancing.
The benefit is in terms of \emph{spectrum reduction ratio},
\be
\rho \triangleq \frac{F^{\textsf{ND}}-F^{\textsf{D2D}}}{F^{\textsf{ND}}}\in [0,1).
\label{equ:spectrum-reduction-ratio}
\ee
The cost is in terms of  \emph{(D2D traffic) overhead ratio},
\be
\eta \triangleq \frac{V^{\textsf{D2D}}}{V^{\textsf{D2D}}+V^{\textsf{BS}}}\in [0,1),
\label{equ:overhead-ratio}
\ee
where $V^{\textsf{D2D}}$ is the volume of all D2D traffic and $V^{\textsf{BS}}$
is the volume of all traffic directly sent by cellular users to BSs.

The spectrum reduction ratio $\rho$ evaluates how much spectrum we can save if we apply D2D load balancing.
The overhead ratio $\eta$ evaluates the percentage of D2D traffic among all traffic.
D2D traffic incurs cost in the sense that any traffic going through D2D links
will consume spectrum and energy of user devices but do not immediately reach any BS.
Overall, the spectrum reduction ratio $\rho$ captures
the benefit of D2D load balancing and hence larger $\rho$ means larger benefit;
the overhead ratio $\eta$ captures the cost of D2D load balancing
and hence smaller $\eta$ means smaller cost.
In the following, we will discuss how to obtain $F^{\textsf{ND}}$ in Sec. \ref{sec:optimal_no_d2d}
and $F^\textsf{D2D}$ in Sec. \ref{sec:optimal_d2d}.
Then we will show the theoretical upper bounds for $\rho$ and $\eta$ in Sec. \ref{sec:theoretical_results}.

\section{Optimal Solution without D2D} \label{sec:optimal_no_d2d}
In this section, we describe how to compute the minimum spectrum
without D2D, i.e., $F^{\textsf{ND}}$. Since there are no D2D links,
we can calculate the required minimum spectrum for each BS separately.
Let us denote $F_b^{\textsf{ND}}$ as the minimum spectrum
of BS $b$ to deliver all its own traffic demands, i.e., $\mathcal{J}_b$.
Then the total minimum spectrum without D2D is\footnote{Here for simplicity, we assume that
all BSs use orthogonal spectrum.
We discuss how to extend our results
to the practical case of spectrum reuse in Sec.~\ref{sec:towards_spectrum_reduction}.}
$
F^{\textsf{ND}} = \sum_{b \in \mathcal{B}} F_b^{\textsf{ND}}.
$

\subsection{Problem Formulation}
For each BS $b \in \mathcal{B}$, we formulate the problem of
minimizing the spectrum to deliver all demands in cell $b$ without D2D,
named as $\textsf{Min-Spectrum-ND}_b$,
\bse
\bee
& \min_{x^{j}_{u_j,b}(t), \gamma_b(t), F_b \in \mathbb{R}^+}  \quad F_b  \label{equ:nd_up_peak_obj}\\
\text{s.t.}
& \quad \sum_{t=s_j}^{e_j} x^{j}_{u_j,b}(t)R_{u_j,b} = r_j, \forall j \in \mathcal{J}_b
\label{equ:nd_up_traffic_cons1} \\
& \quad \sum_{j \in \mathcal{J}_b: t \in [s_j,e_j]} x^{j}_{u_j,b}(t) = \gamma_b(t), \forall t \in [T]
\label{equ:nd_up_peak_cons1} \\
& \quad \gamma_b(t) \le F_b,
\forall t \in [T]
\label{equ:nd_up_peak_cons2} \\
& \quad x^{j}_{u_j,b}(t) \ge 0, \forall j \in \mathcal{J}_b, t \in [s_j, e_j]
\label{equ:nd_up_traffic_cons2}
\eee
\ese
where $x^{j}_{u_j,b}(t)$ is the allocated spectrum (unit: Hz)
for transmitting demand $j$ from user $u_j$ to BS $b$ at slot $t$,
the auxiliary variable $\gamma_b(t)$ is the total used spectrum from users to BS $b$ at slot $t$,
and $F_b$ is the allocated (peak) spectrum to BS $b$,

Our objective is to minimize the total allocated spectrum of BS $b$, as shown in \eqref{equ:nd_up_peak_obj}.
Without D2D, users can only be served by its own BS.
Equation \eqref{equ:nd_up_traffic_cons1} shows the volume requirement for any
traffic demand $j$, i.e., the total traffic volume $r_j$ needs to be delivered from user $u_j$ to BS $b$ during its lifetime.
Equation \eqref{equ:nd_up_peak_cons1} depicts the total needed spectrum of cell $b$ (i.e., $\gamma_b(t)$) in slot $t$,
which is the summation of allocated spectrum for all active jobs in slot $t$.
Inequality \eqref{equ:nd_up_peak_cons2} shows that the total needed spectrum of cell $b$ in any slot $t$
cannot exceed the total allocated spectrum of BS $b$.
Finally, inequality \eqref{equ:nd_up_traffic_cons2} means that the allocated spectrum for a job in any slot is non-negative.

Let us denote $d_{\max} \triangleq \max_{j \in \mathcal{J}} (e_j-s_j+1)$ as
the maximum delay among all demands. Then the number of variables in
$\textsf{Min-Spectrum-ND}_b$ is $O(|\mathcal{J}_b| \cdot d_{\max}+T)$
and the number of constraints in $\textsf{Min-Spectrum-ND}_b$ is also $O(|\mathcal{J}_b| \cdot d_{\max}+T)$.

\subsection{Characterizing the Optimal Solution}
To solve $\textsf{Min-Spectrum-ND}_b$, we can use standard linear programming (LP) solvers.
However, LP solvers cannot exploit the structure of this problem. We next propose a combinatorial algorithm that exploits the problem structure and
achieves lower complexity than general LP algorithms.

We note that $\textsf{Min-Spectrum-ND}_b$ resembles a uniprocessor scheduling problem for preemptive tasks
with hard deadlines \cite{Buttazzo97}. Indeed, we can attach each task $j \in \mathcal{J}_b$ with an
arrival time $s_j$ and a hard deadline $e_j$ and the requested service time
$\frac{r_j}{R_{u_j,b}}$. Then for a given amount of allocated spectrum $F_b$
(which resembles the maximum speed of the processor), we can use
the earliest-deadline-first (EDF) scheduling algorithm  \cite{EDF73} to check its feasibility. Since we
can easily get an upper bound for the minimum spectrum,
we can use binary search to find the minimum spectrum $F_b^{\textsf{ND}}$,
supported by the EDF feasibility-check subroutine.

More interestingly, we can even get a semi-closed form for $F_b^{\textsf{ND}}$, inspired by \cite[Theorem 1]{YDS95}.
Specifically, let us define the \emph{intensity} \cite{YDS95} of an interval $I = [z,z']$
to be
\be
g_b(I) \triangleq \frac{\sum\limits_{j \in \mathcal{A}_b(I)} \frac{r_j}{R_{u_j,b}}}{z'-z + 1}
\label{equ:intensity_nd_b}
\ee
where
$
\mathcal{A}_b(I) \triangleq \{j \in \mathcal{J}_b:
[s_j, e_j] \subset [z,z']\}
$
is the set of all active traffic demands whose lifetime is within the interval $I=[z,z']$.
Then we have the following theorem.

\begin{theorem}
$F_b^{\textsf{ND}} = \max\limits_{I \subset [T]} g_b(I)$.
\label{the:YDS}
\end{theorem}

\begin{IEEEproof}
Since the proof of Theorem 1 was omitted in \cite{YDS95} and
the theorem is not directly mapped to the minimum spectrum problem,
we give a  proof
\ifx \ISTR \undefined
in our technical report \cite{TR} for completeness.
\else
in Appendix \ref{app:proof_YDS} for completeness.
\fi
\end{IEEEproof}

Theorem~\ref{the:YDS} shows that $F_b^{\textsf{ND}}$ is the maximum intensity over all intervals.
To obtain the interval with maximum intensity (and hence $F_b^{\textsf{ND}}$),
we adapt the  algorithm originally developed for solving the job scheduling problem
in \cite{YDS95}, which is called YDS algorithm named after the authors, to our spectrum minimization problem.
The time complexity
of the YDS algorithm is related to the total number of possible intervals. Clearly
the optimal interval can only begin from the generation time of a demand and end
at the deadline of a demand. So the total number of intervals needed to be checked is $O(|\mathcal{J}_b|^2)$.
Thus the time complexity of our adaptive YDS algorithm is $O(|\mathcal{J}_b|^2)$ \cite{YDS95}.
But the complexity of general LP algorithms is $O((|\mathcal{J}_b| \cdot d_{\max}+T)^4L)$
where $L$ is a parameter determined by the coefficients of the LP \cite{khachiyan1980polynomial}.
Thus, our combinatorial algorithm has much lower complexity than general LP algorithms.

\section{Optimal Solution With D2D} \label{sec:optimal_d2d}
In this section, we formulate the optimization problem to compute
the minimum sum spectrum $F^{\textsf{D2D}}$ when D2D communication
is enabled. In this case, since the traffic can be directed
to other BSs via inter-cell D2D links, all BSs are coupled with
each other and need to be considered as a whole. We will first
define the traffic scheduling policy with D2D and then formulate
the problem as an LP.

\subsection{Traffic Scheduling Policy} \label{sec:traffic}
Given traffic demand set $\mathcal{J}$, we need to find a routing policy to
forward each packet to BSs before the deadline, which is the
\emph{traffic scheduling problem}.
Since we should consider the traffic flow in each slot,
we will use the \emph{time-expanded graph} to model the traffic flow over time \cite{Skutella09}.
Specifically, denote $x_{u,v}^{j}(t)$ as the allocated
spectrum (unit: Hz) for link $(u,v)$ at slot $t$ for demand $j \in \mathcal{J}$.
Then the delivered traffic volume from node $u$ to node $v$ at slot $t$ for demand $j$
is $x_{u,v}^{j}(t) R_{u,v}$.
For ease of formulation, we set the self-link rate to be $R_{u,u}=1$.
Then the self-link traffic
i.e., $x_{u,u}^{j}(t)R_{u,u}=x_{u,u}^{j}(t)$,
is the traffic volume stored in node $u$ at slot $t$ for demand $j$.
But the allocated (virtual) spectrum for self-link traffic, i.e., $x_{u,u}^{j}(t)$, will
not contribute to the spectrum requirements of BSs (see \eqref{equ:d2d_peak_cons2} later).
All traffic flows over time are precisely captured by
the time-expanded graph and $x_{u,v}^{j}(t)$.
Then we  define the \emph{traffic scheduling policy} as follows.
\begin{definition}
A {traffic scheduling policy} is the set
$\{x_{u,v}^{j}(t): (u,v) \in \mathcal{E}, j \in \mathcal{J},
t \in [s_j, e_j]\} \cup \{x_{u,u}^{j}(t): u \in \mathcal{V},
j \in \mathcal{J}, t \in [s_j, e_j]\}$
such that
\begin{subequations}
\bee
& \sum_{v\in \text{out}(u_j)}x_{u_j,v}^{j}(s_j)R_{u_j,v}= r_j, \forall j \in \mathcal{J} \label{equ_demand}\\
&\sum_{b\in \mathcal{B}}\sum_{v\in \text{in}(b)}x_{v,b}^{j}(e_j)R_{v,b}= r_j, \forall j \in \mathcal{J} \label{equ_reach}\\
& \sum_{v\in\text{in}(u)}x_{v,u}^{j}(t) R_{v,u}= \sum_{v\in \text{out}(u)}x_{u,v}^{j}(t+1)R_{u,v}, \nnb \\
& \qquad \forall  j \in \mathcal{J}, u \in \mathcal{V}, t \in [s_j, e_j-1]
\label{equ_conservation} \\
& x_{u,v}^{j}(t) \ge 0, \forall (u,v) \in \mathcal{E}, j \in \mathcal{J}, t \in [s_j, e_j]
\label{equ:y_nonnegative} \\
& x_{u,u}^{j}(t) \ge 0, \forall u \in \mathcal{V}, j \in \mathcal{J}, t \in [s_j, e_j]
\label{equ:y_nonnegative_selflink}
\eee
\end{subequations}
where $\text{in}(u)=\{v: (v,u) \in \mathcal{E}\} \cup  \{u\} $
and $\text{out}(u)=\{v:(u,v) \in \mathcal{E}\}  \cup \{u\} $
are the incoming neighbors and outgoing neighbors of
node $u \in \mathcal{V}$ in the time-expanded graph.
\end{definition}

Constraint (\ref{equ_demand}) shows the flow balance in the source node while (\ref{equ_reach})
shows the flow balance in the destination nodes such that
all traffic can reach BSs before their deadlines.
Equality (\ref{equ_conservation}) is the flow conservation constraint
for each intermediate node in the time-expanded graph.
Here we assume that all BSs and all users have enough radios such that
they can simultaneously transmit data to and receive data from multiple BSs (or users).
This is a strong assumption for mobile users because current
mobile devices are not equipped with enough radios. However,
multi-radio mobile devices could be a trend and
there are  substantial research work in multi-radio wireless systems (see a survey in \cite{Si10}
and the references therein). We made this assumption here because
\emph{wireless scheduling problem} for single-radio users is generally
intractable and we want to avoid detracting our attention and focus
on how to characterize the benefit of D2D load balancing and get
a first-order understanding. We  remark that this assumption is also
made in recent work \cite{Zhou20171000Cell} on spectrum reallocation in small-cell cellular networks.

\subsection{Problem Formulation} \label{sec:problem_formulation_D2D}
Then we formulate the problem of computing the minimum total spectrum to serve all demands in all cells with D2D,
named as $\textsf{Min-Spectrum-D2D}$,
\bse \label{equ:d2d-lp}
\bee
& \min_{x^{j}_{u,v}(t), \alpha_b(t),  \beta_b(t), F_b \in \mathbb{R}^+}  \quad \sum_{b \in \mathcal{B}} F_b\\
\text{s.t.} & \quad \eqref{equ_demand}, \eqref{equ_reach},
\eqref{equ_conservation}, \eqref{equ:y_nonnegative}, \eqref{equ:y_nonnegative_selflink} \nnb \\
& \quad \sum_{v \in \mathcal{U}_b} \sum_{j \in \mathcal{J}: t \in [s_j, e_j]} x^{j}_{v,b}(t) = \alpha_b(t),   \forall b \in \mathcal{B}, t \in [T]
\label{equ:d2d_peak_cons1}\\
& \quad \sum_{u \in \mathcal{U}_b} \sum_{v \in \text{in}\left( u \right)
\backslash \left\{ u \right\}} \sum_{j \in \mathcal{J}: t \in [s_j, e_j]} x^{j}_{v,u}(t) = \beta_b(t),  \nnb \\
& \qquad \qquad \forall b \in \mathcal{B}, t \in [T]
\label{equ:d2d_peak_cons2}\\
& \quad \alpha_b(t)+\beta_b(t) \le F_b,\forall b \in \mathcal{B}, t \in [T]
\label{equ:d2d_peak_cons3}
\eee
\ese
where the auxiliary variable $\alpha_b(t)$ is the total used spectrum  from users to BS $b$ at slot $t$,
the auxiliary variable $\beta_b(t)$ is  the total used spectrum  dedicated to all users in BS $b$ at slot $t$,
and $F_b$ is the allocated (peak) spectrum for BS $b$. Note that in our case with D2D load balancing,
a user can adopt \emph{the D2D mode} to transmit to another user via a D2D link
(e.g., $\sum_{j \in \mathcal{J}: t \in [s_j, e_j]} x^{j}_{v,u}(t)$ is the allocated spectrum to the D2D link from user $v$ to user $u$ in slot $t$)
and/or \emph{the cellular mode} to transmit to its BS via a user-to-BS link
(e.g., $\sum_{j \in \mathcal{J}: t \in [s_j, e_j]} x^{j}_{v,b}(t)$ is the allocated spectrum to the user-to-BS link from user $v$ to BS $b$ in slot $t$).
In addition, note that we assume a \emph{receiver-takeover} scheme in the sense that
any traffic will consume spectrum resources  of the receiver's BS.
Equalities \eqref{equ:d2d_peak_cons1} and \eqref{equ:d2d_peak_cons2} show that
BS $b$ is responsible for all traffic dedicated to itself and to its users except self-link (virtual) spectrum (see Sec.~\ref{sec:traffic}).
We also remark that although spectrum sharing is one of the major benefits of D2D communication,
in this work we do not model the spectrum sharing among D2D links and user-to-BS links to simplify the analysis. Later in Sec.~\ref{sec:simulation},
we show that our D2D load balancing scheme can significantly reduce the spectrum requirement even without doing
spectrum sharing among D2D links and user-to-BS links. If we further do spectrum sharing, the
D2D load balancing has more gains.

Given an optimal solution to \textsf{Min-Spectrum-D2D},
we denote $F_b^\textsf{D2D}$ as the allocated spectrum for each BS $b$, and thus the
total spectrum is $F^\textsf{D2D} = \sum_{b \in \mathcal{B}} F_b^\textsf{D2D}.$
The total D2D traffic and total user-to-BS traffic are
\be
V^{\textsf{D2D}} = \sum_{t=1}^{T} \sum_{j \in \mathcal{J}: t \in [s_j, e_j-1]}
\sum_{u \in \mathcal{U}} \sum_{v:v \in \mathcal{U}, (u,v) \in \mathcal{E}} x^{j}_{u,v}(t) {R_{u,v}},
\label{equ:D2D-traffic}
\ee
\be
V^{\textsf{BS}} = \sum_{t=1}^{T} \sum_{j \in \mathcal{J}: t \in [s_j, e_j]}
\sum_{b\in \mathcal{B}} \sum_{u \in \mathcal{U}_b} x^{j}_{u,b}(t) {R_{u,b}},
\label{equ:BS-traffic}
\ee
which are used to calculate the overhead ratio $\eta$ in \eqref{equ:overhead-ratio}.
We further remark that since all traffic demands must reach any BSs, it is easy to see that
the user-to-BS traffic is exactly the total volume of all traffic demands, i.e.,
$
V^{\textsf{BS}} = \sum_{j \in \mathcal{J}} r_j.
$

Given the optimal (minimum) total spectrum, i.e., $F^{\textsf{D2D}}$,
we next minimize the overhead, named \textsf{Min-Overhead}, by solving the following LP\footnote{In other words, minimizing the
total spectrum is our first-priority objective and minimizing the corresponding D2D traffic overhead (without exceeding
the minimum total spectrum) is our second-priority objective.},
\bse \label{equ:overhead-lp}
\bee
& \min_{\substack{x^{j}_{u,v}(t), \alpha_b(t), \\  \beta_b(t), F_b \in \mathbb{R}^+}}  \quad \sum_{t=1}^{T} \sum_{j \in \mathcal{J}: t \in [s_j, e_j-1]}
\sum_{u \in \mathcal{U}} \sum_{ \substack{v:v \in \mathcal{U}, \\ (u,v) \in \mathcal{E}}} x^{j}_{u,v}(t) {R_{u,v}}\\
& \text{s.t.}  \quad \eqref{equ_demand}, \eqref{equ_reach},
\eqref{equ_conservation}, \eqref{equ:y_nonnegative}, \eqref{equ:y_nonnegative_selflink} \nnb \\
& \quad \sum_{v \in \mathcal{U}_b} \sum_{j \in \mathcal{J}: t \in [s_j, e_j]} x^{j}_{v,b}(t) = \alpha_b(t),  \forall b \in \mathcal{B}, t \in [T]
\label{equ:overhead_peak_cons1}\\
& \quad \sum_{u \in \mathcal{U}_b} \sum_{v \in \text{in}\left( u \right)
\backslash \left\{ u \right\}} \sum_{j \in \mathcal{J}: t \in [s_j, e_j]} x^{j}_{v,u}(t) = \beta_b(t),  \nnb \\
& \qquad \qquad  \forall b \in \mathcal{B}, t \in [T]
\label{equ:overhead_peak_cons2}\\
& \quad \alpha_b(t)+\beta_b(t) \le F_b,\forall b \in \mathcal{B}, t \in [T]
\label{equ:overhead_peak_cons3} \\
& \quad \sum_{b \in \mathcal{B}} F_b \le F^{\textsf{D2D}}
\label{equ:overhead_peak_cons4}
\eee
\ese

As compared to $\textsf{Min-Spectrum-D2D}$ in \eqref{equ:d2d-lp},
\textsf{Min-Overhead} in \eqref{equ:overhead-lp} adds a constraint
\eqref{equ:overhead_peak_cons4} for the \emph{given} total spectrum
$F^{\textsf{D2D}}$ and changes the objective to be
the total D2D traffic defined in \eqref{equ:D2D-traffic}.
Note that even though we write \eqref{equ:overhead_peak_cons4}
as an inequality, it must hold as an equality. This is because
$F^{\textsf{D2D}}$ is  the optimal value of $\textsf{Min-Spectrum-D2D}$ in \eqref{equ:d2d-lp}
and any solution in \textsf{Min-Overhead} in \eqref{equ:overhead-lp} is also feasible
to $\textsf{Min-Spectrum-D2D}$ in \eqref{equ:d2d-lp}.

The number of variables in \textsf{Min-Spectrum-D2D}  is
$O(|\mathcal{J}|\cdot|\mathcal{E}|\cdot d_{\max} + |\mathcal{B}| \cdot T)$
and the number of constraints in \textsf{Min-Spectrum-D2D}  is
$O(|\mathcal{J}| \cdot (|\mathcal{V}|+|\mathcal{E}|) \cdot d_{\max} + |\mathcal{B}| \cdot T )$.
The problem \textsf{Min-Overhead} has the same complexity as
 \textsf{Min-Spectrum-D2D}.
Solving the problem, even though it is an LP, incurs high complexity.
We further discuss how to reduce the complexity without loss of optimality
\ifx \ISTR \undefined
in our technical report \cite{TR}.
\else
in Appendix~\ref{sec:time_space_complexity}.
\fi
Even with our optimized LP approach,
later in our simulation in Sec.~\ref{sec:simulation}, we show that we cannot
solve \textsf{Min-Spectrum-D2D} for  practical Smartone network
with off-the-shell servers. Thus, we further propose a heuristic algorithm
to solve \textsf{Min-Spectrum-D2D} with much lower complexity in Sec.~\ref{sec:heuristic}.
We also provide performance guarantee for our heuristic algorithm.
Before that, we show our theoretical results on the spectrum reduction ratio and the overhead ratio
in next section.

\section{Theoretical Results} \label{sec:theoretical_results}
From the two preceding sections, we can compute $F^{\textsf{ND}}$ with
the (adaptive) YDS algorithm (Theorem \ref{the:YDS}) and $F^{\textsf{D2D}}$ by
solving the large-scale LP problem \textsf{Min-Spectrum-D2D} (Sec. \ref{sec:problem_formulation_D2D}).
Hence, numerically we can get the  spectrum reduction and
the overhead ratio. In this section, however, we seek to
derive theoretical upper bounds on both spectrum reduction and overhead ratio.
Such theoretical upper bounds provide insights for the key factors to
achieve large spectrum reduction and thus provide guidance to determine whether
it is worthwhile to implement D2D load balancing scheme in real-world cellular systems.

%

%

\subsection{A Simple Upper Bound for Spectrum Reduction}
We can get a simple upper bound for $F^{\textsf{D2D}}$ by assuming no cost for D2D communication in the sense
that any D2D communication will not consume bandwidth and will not incur delays. Then we can
construct a virtual grand BS and all users $\mathcal{U}$ are in this BS. Then the system becomes similar
to the case without D2D. We can apply the YDS algorithm
to compute the minimum peak traffic, which is a lower bound
for $F^{\textsf{D2D}}$, i.e.,
$
\underline{F}^{\textsf{D2D}}=  \max_{I \subset [T]} g(I),
$
where
\be
g(I) = \frac{\sum\limits_{j \in \mathcal{A}(I)} \frac{r_j}{R_{\max}}}{z'-z + 1}.
\label{equ:intensity_d2d}
\ee
Here in \eqref{equ:intensity_d2d},
$
\mathcal{A}(I) = \{j \in \mathcal{J}: [s_j, e_j] \subset [z,z']\}
$
is the set of all active traffic demands whose lifetime is within the interval $I=[z,z']$
and $R_{\max} = \max_{s \in \mathcal{U}} R_{s,b_s}$ is the best user-to-BS link. Then we have the following theorem.


\begin{theorem}
$\rho \le \frac{F^{\textsf{ND}}-\underline{F}^{\textsf{D2D}}}{F^{\textsf{ND}}}$.
\label{the:trivial_upper_bound}
\end{theorem}
\begin{IEEEproof}
\ifx \ISTR \undefined
Please see our technical report \cite{TR}.
\else
Please see Appendix~\ref{app:proof_trivial_upper_bound}.
\fi
\end{IEEEproof}

Note that both $\underline{F}^{\textsf{D2D}}$ and $F^{\textsf{ND}}$ can be computed by the YDS algorithm, much
easier than solving the large-scale LP \textsf{Min-Spectrum-D2D}. Therefore, numerically we can get
a quick understanding of the maximum benefit that can be achieved by D2D load balancing.

\subsection{A General Upper Bound for Spectrum Reduction} \label{sec:a_general_upper_bound}
We next describe another general upper
bound for any arbitrary topology and any arbitrary traffic demand set. We will begin with
some preliminary notations.

\begin{figure*}
\begin{minipage}[c]{0.32\linewidth}
\captionof{table}{Discrepancy Notations.} \label{tab:Discrepancy Notations}
\begin{tabular}{|l|}
\hline
\scriptsize{$r_s = \max_{v: (s,v) \in \mathcal{E}, v \in \mathcal{U}_{b_s}} \frac{R_{s,v}}{R_{s,b_s}},
 \forall s \in \mathcal{U}$}
\\%
\scriptsize{$\tilde{r}_s^b = \max_{v: (s,v) \in \mathcal{E}, v \in \mathcal{U}_{b}} \frac{R_{s,v}}{R_{s,b_s}},
 \forall s \in \mathcal{U}, b \in \mathcal{B}$}
\\
\scriptsize{$r_b = \max_{s \in \mathcal{U}_b} r_s, \forall b \in \mathcal{B}$}
\\
\scriptsize{$\tilde{r}_{b,b'} = \max_{s \in \mathcal{U}_b} \tilde{r}_s^{b'}, \forall b \in \mathcal{B}, b' \in \mathcal{B}$}
\\
\scriptsize{$r = \max_{b \in \mathcal{B}}{r_b}$,  $\tilde{r} = \max_{(b,b') \in \mathcal{E}^{\textsf{D2D}}} \tilde{r}_{b,b'}$}
\\
\hline
\end{tabular}
\end{minipage}
\hfill
\begin{minipage}[c]{0.3\linewidth}
  \centering
  \includegraphics[width=\linewidth]{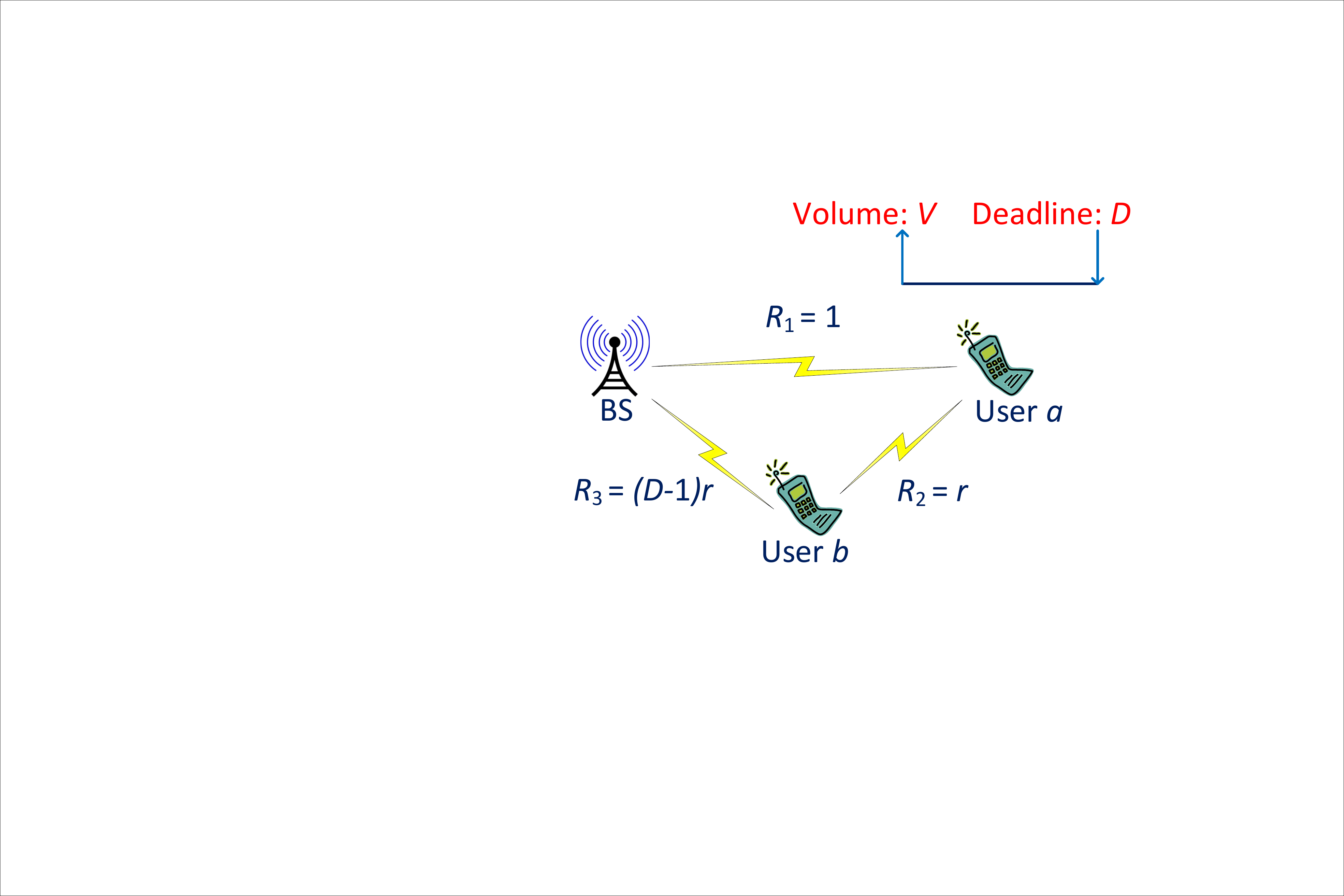}\\
  \caption{The benefit of intra-cell D2D communications.}
  \label{fig:intra_cell_benefit}
\end{minipage}
\hfill
\begin{minipage}[c]{0.3\linewidth}
\centering
\includegraphics[width=\linewidth]{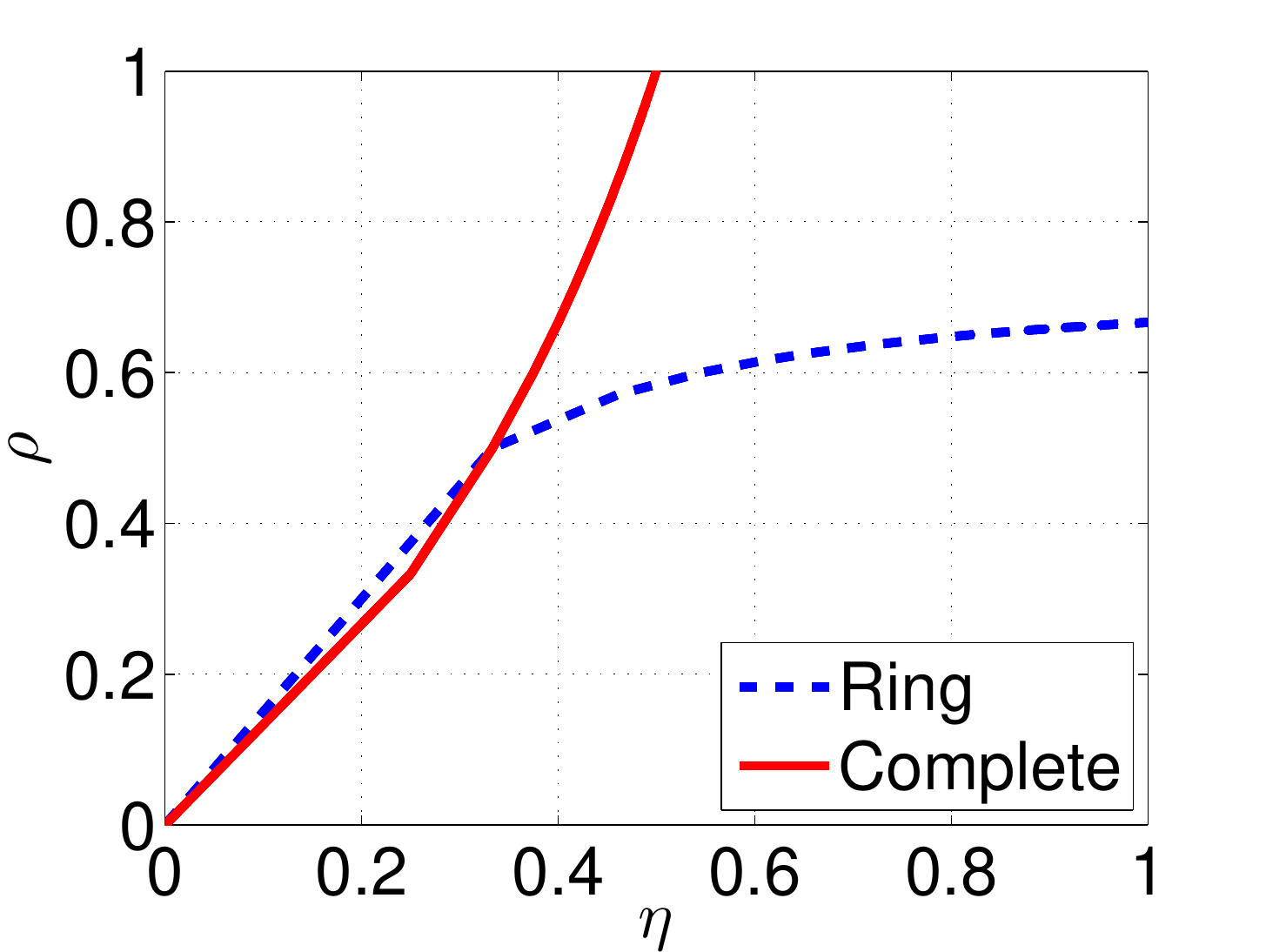}
\caption{\label{fig:tradeoff_ring_complete}  Tradeoff between $\rho$ and $\eta$.}
\end{minipage}
\end{figure*}

We first define some preliminary notations. Let $N=|\mathcal{B}|$ be the number of BSs and we define
a directed \emph{D2D communication graph} $\mathcal{G}^{\textsf{D2D}}= (\mathcal{B}, \mathcal{E}^{\textsf{D2D}})$
where the vertex set is the BS set $\mathcal{B}$ and $(b,b') \in \mathcal{E}^{\textsf{D2D}}$
if there  exists at least one inter-cell D2D link from user $u \in \mathcal{U}_b$ in BS $b \in \mathcal{B}$
to user $v \in \mathcal{U}_{b'}$ in BS $b' \in \mathcal{B}$.
Denote $\delta^-_b$ as the in-degree of BS $b$ in the graph $\mathcal{G}^{\textsf{D2D}}$
and define the maximum in-degree of the graph $\mathcal{G}^{\textsf{D2D}}$ as $\Delta^- = \max_{b \in \mathcal{B}} \delta^-_b$.
In addition, we define some notations in Tab. \ref{tab:Discrepancy Notations}
to capture the discrepancy of D2D links and non-D2D links for users and BSs.
Note that these definitions will be used thoroughly
\ifx \ISTR \undefined
in our technical report \cite{TR} to prove {Theorem} \ref{the:ratio_bound}.
\else
in Appendix \ref{app:proof_upper_bound} to prove {Theorem} \ref{the:ratio_bound}.
\fi

%

Now we have the following theorem.

\begin{theorem}
\label{the:ratio_bound}
For an arbitrary network topology $\mathcal{G}$ associated with a D2D communication
graph $\mathcal{G}^{\textsf{D2D}}= (\mathcal{B}, \mathcal{E}^{\textsf{D2D}})$ and an arbitrary
traffic demand set, the  spectrum reduction is upper bounded by
\be
\rho \le \frac{\max\{r,1\} + \tilde{r} \Delta^--1}{\max\{r,1\} + \tilde{r} \Delta^-}.
\ee
\end{theorem}
\begin{IEEEproof}
\ifx \ISTR \undefined
Please see Appendix \ref{app:proof_upper_bound}.
\else
Please see  our technical report \cite{TR}.
\fi
\end{IEEEproof}

Based on this upper bound, we observe that the benefit of D2D load
balancing comes from two parts: intra-cell D2D and inter-cell D2D.
More interestingly, we can obtain the individual benefit of intra-cell
D2D and inter-cell D2D separately, as shown in the following Corollaries \ref{cor:intra_cell_benefit}
and \ref{cor:inter_cell_benefit}. One can go through the proof for Theorem \ref{the:ratio_bound}
by disabling inter-cell or intra-cell D2D communication and get the proof of these
two corollaries.

\begin{corollary} \label{cor:intra_cell_benefit}
If only intra-cell D2D communication is enabled, the  spectrum reduction is upper bounded by
\be
\rho \le \frac{\max\{r,1\} -1}{\max\{r,1\}}.
\ee
\end{corollary}

This upper bound is quite intuitive. When $r \le 1$, then for any user $s$, there does
not exist any intra-cell D2D link with better link quality than its direct link to BS $b_s$.
Therefore, using the user-to-BS link is always the optimal choice. Thus the spectrum
reduction is 0. When $r > 1$, larger $r$ means more advantages for intra-cell D2D links
over the user-to-BS links. Therefore, D2D can exploit more benefit.

Moreover, this upper bound can be achieved by the simple example in Fig. \ref{fig:intra_cell_benefit}.
Suppose that user $a$ generates one traffic demand with volume $V$ and delay $D \ge 2$ at slot 1.
Suppose link rates $R_1=1, R_2=r, R_3=(D-1)r$.
Then without intra-cell D2D, the (peak) spectrum requirement is $F_1=\frac{V}{D}$. With intra-cell
D2D, user $a$ transmits $\frac{V}{D-1}$ traffic to user b from slot
1 to slot $D-1$ and then user $b$ transmits all $V$ traffic to BS at slot $D$.
The (peak) spectrum requirement is $F_2 = \max\{\frac{V}{(D-1)R_2}, \frac{V}{R_3}\} = \frac{V}{(D-1)r}$.
Then the spectrum reduction is
\be
\frac{F_1-F_2}{F_1} = 1 - \frac{\frac{V}{(D-1)r}}{\frac{V}{D}} \to \frac{r-1}{r},
\text{as} \; D \to \infty.
\ee

The benefit of intra-cell D2D communication is widely studied (see \cite{Doppler09} \cite{Foder12}).
However, in this paper, we mainly focus on the benefit of inter-cell D2D load balancing. Indeed,
in our simulation settings in Sec. \ref{sec:simulation}, the intra-cell D2D brings negligible benefit.

\begin{corollary} \label{cor:inter_cell_benefit}
If only inter-cell D2D communication is enabled, the  spectrum reduction is upper bounded by
$
\rho \le \frac{\tilde{r} \Delta^-}{1 + \tilde{r} \Delta^-}.
$
\end{corollary}

The intuition behind the parameter $\tilde{r}$ is similar to the effect of
parameter $r$ in the intra-cell D2D case. In what follows, we will
only discuss the effect of parameter $\Delta^-$, which actually
reveals the insight of our advocated D2D load balancing scheme.
Now suppose that all the links have the
same quality and \emph{w.l.o.g.} let $R_{u,v}=1, \forall (u,v) \in \mathcal{E}$. Then
$r = \tilde{r} =1$, meaning that no intra-cell D2D benefit exists. And the benefit
of inter-cell D2D is reduced to the following upper bound
\be
\rho \le \frac{\Delta^-}{1+\Delta^-}.
\label{equ:bound_delta}
\ee

The rationale to understand this upper bound is as follows.
On a high level of understanding,
the main idea for load balancing is traffic aggregation.
If each BS can aggregate more traffic from other BSs, it can exploit more
statistical multiplexing gains to serve more traffic with the same amount of spectrum.
Since the in-degree for each BS indeed measures its capacity of traffic aggregation,
it is not surprising that the upper bound for $\rho$ is related to maximum in-degree $\Delta^-$.

To evaluate how good the upper bound in \eqref{equ:bound_delta} is, two natural questions can
be asked. The first is: \emph{Is this upper bound tight}?
Another observation is that if we want to achieve unbounded benefit,
i.e., $\rho \to 1$,
it is necessary to let $\frac{\Delta^-}{\Delta^-+1} \to 1$,
which means that $\Delta^- \to \infty$.
Then the second question is:
\emph{Can $\rho$ indeed approach 100\% as  $\Delta^- \to \infty$}?

In the rest of this subsection, we will answer these two questions
by  constructing a specified network and traffic demand set.
Specifically, we consider $N=|\mathcal{B}|$ BSs each serving one user only. To facilitate analysis, let
$b_i$ be the $i$-th BS and $u_i$ be the user in BS $i$, for
all $i\in[N]$. We consider a \emph{singleton-decoupled} traffic
demand set as follows. Each user has one and only one traffic
demand with the same volume $V$ and the same delay $D \ge 2$.
Let $T=ND$ and the traffic generation time of user $i$ be slot $D(i-1)+1$.
Therefore, the lifetime of user $u_i$'s traffic demand is $[D(i-1)+1,Di]$,
during which there are no other demands.

Under such settings, we will vary the user-connection pattern such that
the D2D communication graph is different.
Specifically, we will prove that this upper bound is asymptotically
tight in the ring topology for $\Delta^-=2$ in Fact \ref{fact:bs_ring},
and $\rho \to$ 100\% in the complete topology
as the number of BSs $N \to \infty$ in Fact \ref{fact:bs_complete}.
Moreover, we will also discuss the overhead ratio for these
two special topologies.

\begin{fact} \label{fact:bs_ring}
If $N=2D-1$ and the D2D communication graph
forms a bidirectional ring graph, then
there exists a traffic scheduling policy
such that the  spectrum reduction is
\be
\rho =\frac{2(D-1)}{3D-2} \to \frac{2}{3} = \frac{\Delta^-}{\Delta^-+1},
\text{as} \; {D \to \infty}.
\ee
Besides, the overhead ratio in this case is
\be
\eta = \frac{D(D-1)}{D^2+2D-2}.
\ee
\end{fact}
\begin{IEEEproof}
\ifx \ISTR \undefined
Please see  our technical report \cite{TR}.
\else
Please see Appendix \ref{app:proof_ring}.
\fi
\end{IEEEproof}
\begin{fact} \label{fact:bs_complete}
If the D2D communication graph forms a bidirectional complete graph,
then there exists a traffic scheduling policy such that the spectrum reduction is
\be
\rho = \frac{N-1}{N+1} \to 100\%, \text{as} \; N \to \infty.
\ee
Besides, the overhead ratio in this case is
\be
\eta = \frac{N-1}{2N}.
\ee
\end{fact}
\begin{IEEEproof}
\ifx \ISTR \undefined
Please see  our technical report \cite{TR}.
\else
Please see Appendix \ref{app:proof_complete}.
\fi
\end{IEEEproof}

\textbf{Remark:}
(i) Fact \ref{fact:bs_ring} shows the tightness of the upper bound in \eqref{equ:bound_delta}
for the ring-graph topology when $\Delta^-=2$.
(ii) { Fact \ref{fact:bs_complete} shows that $\rho$ can indeed approach $100\%$,
implying that in the best case, $\rho$ goes to $100\%$.
This  gives us strong motivation to investigate D2D load balancing scheme both theoretically and practically.}
(iii) For the complete-graph topology, the upper bound $\frac{\Delta^-}{\Delta^-+1}$ is not tight.
Indeed, since $\Delta^- =N-1$ in the complete-graph topology,
we have
\bee
\frac{\Delta^-}{\Delta^-+1} = \frac{N-1}{N} > \frac{N-1}{N+1}.
\eee
(iv) Let us revisit the toy example in Fig. \ref{fig:example}
which forms a complete-graph topology with $N=2$. It verifies the spectrum reduction
and overhead ratio in Fact \ref{fact:bs_complete},
i.e., $\rho = \frac{1}{3} = \frac{N-1}{N+1}$ and $\eta = \frac{1}{4} = \frac{N-1}{2N}$.
(v) We also highlight the tradeoff between the benefit $\rho$ and the cost $\eta$,
as illustrated in Fig. \ref{fig:tradeoff_ring_complete}.
Furthermore, Fig. \ref{fig:tradeoff_ring_complete} shows that
the complete-graph topology outperforms the ring-graph topology asymptotically
because $\rho \to \frac{2}{3}$ and $\eta \to 1 $ for the ring-graph topology but
$\rho \to 1 > \frac{2}{3}$ (larger benefit) and $\eta \to \frac{1}{2} < 1$ (smaller cost) for the complete-graph topology.

\subsection{An Upper Bound for Overhead Ratio}
Previously we study upper bounds for the spectrum reduction. Now
we instead propose an upper bound for overhead ratio. Recall that
$d_{\max}$ is the maximum demand delay. We then have the following result.

\begin{theorem} \label{the:overhead-upper-bound}
$\eta \le \frac{d_{\max}-1}{d_{\max}}$.
\end{theorem}
\begin{IEEEproof}
\ifx \ISTR \undefined
Please our technical report \cite{TR}.
\else
Please see Appendix~\ref{app:proof-of-overhead-upper-bound}.
\fi
\end{IEEEproof}

The upper bound in Theorem~\ref{the:overhead-upper-bound} increases when the maximum
demand delay $d_{\max}$ increases. This is reasonable because a traffic demand can travel more
D2D links (and thus incurs more D2D traffic overhead) if its delay is large.
For our toy example in Fig.~\ref{fig:example}, we have $d_{\max}=2$ and thus
the upper bound for the overhead ratio is $\frac{d_{\max}-1}{d_{\max}}=50\%$, which
is in line with our actual overhead ratio 25\%.

\section{A Low-Complexity Heuristic Algorithm for \textsf{Min-Spectrum-D2D}}  \label{sec:heuristic}
Our proposed LP formulation for \textsf{Min-Spectrum-D2D} has high complexity
due to the size of input traffic demand and cellular network.
To reduce the complexity, in this section, we propose a heuristic algorithm which
can significantly reduce the number of traffic demands that is needed to be considered. Moreover,
our algorithm has a parameter (which is $\lambda$ defined shortly)
such that we can balance the complexity and
the performance.

Our proposed algorithm has three steps.
\par\noindent\rule{0.5\textwidth}{0.4pt}

\emph{Step I.}
We solve \textsf{Min-Spectrum-ND}$_b$ for each BS $b \in \mathcal{B}$, and get
the optimal solution $\{x^{j}_{u_j,b}(t), \gamma_b(t), F_b\}$.

\emph{Step II.}
For each BS $b$ with the spectrum profile $\gamma_b(t)$, we consider the following set,
\be
T_b(\lambda) \triangleq \{t \in [T]: \gamma_b(t) > \lambda F_b \},
\label{equ:T-b-lambda}
\ee
where parameter $\lambda \in [0,1]$ controls the split level.
Now we divide all cell-$b$ traffic demands $\mathcal{J}_b$ into two demand sets
\be
\resizebox{0.89\linewidth}{!}{$\mathcal{J}_b^{\textsf{D2D}}(\lambda) \triangleq \{j \in \mathcal{J}_b:
 \exists t \in [s_j, e_j] \cap T_b(\lambda) \text{ s.t. } x_{u_j,b}^{j}(t) > 0 \},$}
\label{equ:D2D-demand-set}
\ee
and
\be
\resizebox{0.89\linewidth}{!}{$\mathcal{J}_b^{\textsf{ND}}(\lambda) \triangleq \{j \in \mathcal{J}_b: x_{u_j,b}^{j}(t) = 0, \forall t \in [s_j, e_j] \cap T_b(\lambda)\}.$}
\label{equ:ND-demand-set}
\ee

For all traffic demand in $\mathcal{J}_b^{\textsf{ND}}(\lambda)$, we  schedule them according to $\{x^{j}_{u_j,b}(t)\}$ without D2D,
which results in at most $\gamma_b(t)$ spectrum requirement for BS $b$ at slot $t$.
Note that no demand in $\mathcal{J}_b^{\textsf{ND}}(\lambda)$
is served in slot set $T_b(\lambda)$. We thud denote $\tilde{\gamma}_b(t)$ as
the already allocated spectrum spectrum for demand set $\mathcal{J}_b^{\textsf{ND}}(\lambda)$ for BS $b$ at slot $b$,
which satisfies $\tilde{\gamma}_b(t) \le \gamma_b(t)$ when $t \notin T_b(\lambda)$
and $\tilde{\gamma}_b(t) =0$ when $t \in T_b(\lambda)$.

\emph{Step III.}  We solve the D2D load balancing problem with traffic demands
$\mathcal{J}^{\textsf{D2D}}(\lambda) \triangleq \{\mathcal{J}_b^{\textsf{D2D}}(\lambda): b \in \mathcal{B}\}$,
according to the following LP, which adaptes \textsf{Min-Spectrum-D2D} in \eqref{equ:d2d-lp} by considering the already allocated spectrum $\{\tilde{\gamma}_b(t)\}$,
\bse \label{equ:d2d-lp-heuristic}
\bee
& \min_{x^{j}_{u,v}(t), \alpha_b(t),  \beta_b(t), F_b \in \mathbb{R}^+}  \quad \sum_{b \in \mathcal{B}} F_b \\
\text{s.t.} & \quad \eqref{equ_demand}, \eqref{equ_reach},
\eqref{equ_conservation}, \eqref{equ:y_nonnegative}, \eqref{equ:y_nonnegative_selflink} \nnb \\
& \quad \sum_{v \in \mathcal{U}_b} \sum_{j \in \mathcal{J}^{\textsf{D2D}}(\lambda): t \in [s_j, e_j]} x^{j}_{v,b}(t) = \alpha_b(t),  \forall b \in \mathcal{B}, t \in [T]
\label{equ:d2d_peak_cons1-heuristic}\\
& \quad \sum_{u \in \mathcal{U}_b} \sum_{v \in \text{in}\left( u \right)
\backslash \left\{ u \right\}} \sum_{j \in \mathcal{J}^{\textsf{D2D}}(\lambda): t \in [s_j, e_j]} x^{j}_{v,u}(t) = \beta_b(t),  \nnb \\
& \qquad \qquad \forall b \in \mathcal{B}, t \in [T]
\label{equ:d2d_peak_cons2-heuristic}\\
& \quad \alpha_b(t)+\beta_b(t) + \tilde{\gamma}_b(t) \le F_b,\forall b \in \mathcal{B}, t \in [T]
\label{equ:d2d_peak_cons3-heuristic}
\eee
\ese

Similar to the overhead minimization problem \textsf{Min-Overhead} in \eqref{equ:overhead-lp},
given the optimal spectrum requirement of \eqref{equ:d2d-lp-heuristic}, denoted as, $F^{\textsf{Heuristic}}(\lambda)$,
we next minimize the overhead by solving the following LP,
\bse \label{equ:d2d-lp-heuristic-overhead}
\bee
& \min_{\substack{x^{j}_{u,v}(t), \alpha_b(t), \\ \beta_b(t), F_b \in \mathbb{R}^+}}  \quad  \sum_{t=1}^{T} \sum_{j \in \mathcal{J}^{\textsf{D2D}}: t \in [s_j, e_j-1]}
\sum_{u \in \mathcal{U}} \sum_{\substack{v:v \in \mathcal{U}, \\ (u,v) \in \mathcal{E}}} x^{j}_{u,v}(t) {R_{u,v}} \\
\text{s.t.} & \quad \eqref{equ_demand}, \eqref{equ_reach},
\eqref{equ_conservation}, \eqref{equ:y_nonnegative}, \eqref{equ:y_nonnegative_selflink} \nnb \\
& \quad \sum_{v \in \mathcal{U}_b} \sum_{j \in \mathcal{J}^{\textsf{D2D}}(\lambda): t \in [s_j, e_j]} x^{j}_{v,b}(t) = \alpha_b(t),  \forall b \in \mathcal{B}, t \in [T]
\label{equ:d2d_peak_cons1-heuristic-overhead}\\
& \quad \sum_{u \in \mathcal{U}_b} \sum_{v \in \text{in}\left( u \right)
\backslash \left\{ u \right\}} \sum_{j \in \mathcal{J}^{\textsf{D2D}}(\lambda): t \in [s_j, e_j]} x^{j}_{v,u}(t) = \beta_b(t),  \nnb \\
& \qquad \qquad \forall b \in \mathcal{B}, t \in [T]
\label{equ:d2d_peak_cons2-heuristic-overhead}\\
& \quad \alpha_b(t)+\beta_b(t) + \tilde{\gamma}_b(t) \le F_b,  \forall b \in \mathcal{B}, t \in [T]
\label{equ:d2d_peak_cons3-heuristic-overhead} \\
& \quad \sum_{b \in \mathcal{B}} F_b \le F^{\textsf{Heuristic}}(\lambda) \label{equ:sum-spectrum-requirement-overhead}
\eee
\ese

\par\noindent\rule{0.5\textwidth}{0.4pt}

Note that in \eqref{equ:d2d-lp-heuristic}/\eqref{equ:d2d-lp-heuristic-overhead}, all variables, $x^{j}_{u,v}(t), \alpha_b(t),  \beta_b(t), F_b$, have
the same meanings of those in \eqref{equ:d2d-lp}/\eqref{equ:overhead-lp}. There are two differences between \eqref{equ:d2d-lp-heuristic}/\eqref{equ:d2d-lp-heuristic-overhead}
and \eqref{equ:d2d-lp}/\eqref{equ:overhead-lp}. First, the traffic demand set in \eqref{equ:d2d-lp-heuristic}/\eqref{equ:d2d-lp-heuristic-overhead} is
$\mathcal{J}^{\textsf{D2D}}(\lambda)$ while that in \eqref{equ:d2d-lp}/\eqref{equ:overhead-lp} is $\mathcal{J}$.
Likewise, the traffic scheduling policy characterized by \eqref{equ_demand}, \eqref{equ_reach},
\eqref{equ_conservation}, \eqref{equ:y_nonnegative}, \eqref{equ:y_nonnegative_selflink} in \eqref{equ:d2d-lp-heuristic}/\eqref{equ:d2d-lp-heuristic-overhead}  is  for the traffic demand set
$\mathcal{J}^{\textsf{D2D}}(\lambda)$ while that in \eqref{equ:d2d-lp}/\eqref{equ:overhead-lp} is for the traffic demand set $\mathcal{J}$.
Second, constraint \eqref{equ:d2d_peak_cons3-heuristic}/\eqref{equ:d2d_peak_cons3-heuristic-overhead}
is different from constraint \eqref{equ:d2d_peak_cons3}/\eqref{equ:overhead_peak_cons3} in that \eqref{equ:d2d_peak_cons3-heuristic}/\eqref{equ:d2d_peak_cons3-heuristic-overhead} considers the already allocated spectrum $\{\tilde{\gamma}_b(t)\}$.
Namely, the spectrum requirement for BS $b$ at slot $t$ includes the already allocated spectrum
$\tilde{\gamma}_b(t)$ to serve the traffic demand $\mathcal{J}_b^{\textsf{ND}}$
and the new allocated spectrum $(\alpha_b(t)+\beta_b(t))$ to serve the traffic demand
$\mathcal{J}_b^{\textsf{D2D}}(\lambda)$.

Obviously, if the number of traffic demand in $\mathcal{J}^{\textsf{D2D}}(\lambda)$ is much less than the total number of traffic demands in $\mathcal{J}$, which is indeed the case according to our empirical study in Sec.~\ref{sec:simulation},
we can significantly reduce the number of variables and constraints in
\eqref{equ:d2d-lp-heuristic}/\eqref{equ:d2d-lp-heuristic-overhead}
in Step III as compared to the LP problem \textsf{Min-Spectrum-D2D}/\textsf{Min-Overhead} in \eqref{equ:d2d-lp}/\eqref{equ:overhead-lp}.
After these three steps, the total spectrum is given by the objective value of \eqref{equ:d2d-lp-heuristic}
and the corresponding overhead is given by the objective value of \eqref{equ:d2d-lp-heuristic-overhead}.
An example of our heuristic algorithm is shown
\ifx \ISTR \undefined
in our technical report \cite{TR}.
\else
in Appendix~\ref{app:an-example-for-heuristic}.
\fi

We denote the spectrum reduction of our heuristic algorithm as
\be
\rho^{\textsf{Heuristic}}(\lambda) \triangleq \frac{F^{\textsf{ND}}-F^{\textsf{Heuristic}}(\lambda)}{F^{\textsf{ND}}}.
\ee
Similarly, we denote $\eta^{\textsf{Heuristic}}(\lambda)$ as the overhead ratio of our heuristic algorithm.
We next show that the performance guarantee of our heuristic algorithm.

First, for the spectrum we reduction, we have,
\begin{theorem} \label{thm:performance-heuristic}
$(1-\lambda) \rho \le \rho^{\textsf{Heuristic}}(\lambda) \le \rho.$
\end{theorem}
\begin{IEEEproof}
\ifx \ISTR \undefined
Please see Appendix~\ref{app:proof-performance-heuristic}.
\else
Please see our technical report \cite{TR}.
\fi
\end{IEEEproof}

Theorem \ref{thm:performance-heuristic} shows that when $\lambda=0$, we have
$\rho^{\textsf{Heuristic}}(0) = \rho$. This is because when $\lambda=0$, we have
$\mathcal{J}^{\textsf{D2D}}(0)=\mathcal{J}$, i.e., all demands  participate
in D2D load balancing in our heuristic algorithm when $\lambda=0$ and thus
the objective value of \eqref{equ:d2d-lp-heuristic} when $\lambda=0$ is exactly $F^\textsf{D2D}$.
When $\lambda=1$, since $\mathcal{J}^{\textsf{D2D}}(1)=\emptyset$, all traffic demands are served locally without D2D and
therefore the objective value of \eqref{equ:d2d-lp-heuristic} when $\lambda=1$ is exactly $F^\textsf{ND}$.
Thus, the lower bound $(1-\lambda) \rho = 0$  is tight.
Further, the lower bound $(1-\lambda) \rho$,
decreases as $\lambda$ increases, but the computational complexity
decreases as $\lambda$ increases. Thus, this lower bound illustrates
the tradeoff between the performance and the complexity of our heuristic algorithm.

Second, we give an upper bound for the overhead ratio\footnote{Recall that $d_{\max}$ is the maximum demand delay.}.
\begin{theorem} \label{the:overhead-upper-bound-heuristic}
$\eta^{\textsf{Heuristic}}(\lambda) \le \frac{ (d_{\max}-1) \sum\limits_{j \in \mathcal{J}^{\textsf{D2D}  }(\lambda)} r_j }{
(d_{\max}-1) \sum\limits_{j \in \mathcal{J}^{\textsf{D2D}  }(\lambda)} r_j + \sum\limits_{j \in \mathcal{J}} r_j} \le \frac{d_{\max}-1}{d_{\max}}$.
\end{theorem}
\begin{IEEEproof}
\ifx \ISTR \undefined
Please see our technical report \cite{TR}.
\else
Please see Appendix \ref{app:proof-of-overhead-upper-bound-heuristic}.
\fi
\end{IEEEproof}
We can see that the upper bound of the overhead ratio is 0 when $\lambda=1$ because $\mathcal{J}^{\textsf{D2D}}(1)=\emptyset$, i.e.,
all traffic demands are served locally without D2D. Moreover, when $\lambda$ increases, the upper bound decreases because
less traffic demands participate in D2D load balancing.

Overall, our heuristic algorithm
reduce the complexity of our global LP approach and has performance guarantee.
Moreover, our proposed heuristic algorithm has a controllable parameter $\lambda$ to
balance the benefit in terms of spectrum reduction, the cost in terms of overhead ratio, and
the computational complexity for our D2D load balancing scheme.

\section{Towards Spectrum Reduction with Frequency Reuse} \label{sec:towards_spectrum_reduction}
In this paper, we use the sum spectrum to describe how many resources are needed to
serve all users' traffic demands in cellular networks.
This may not directly reflect the total required spectrum for cellular operators, because
the same spectrum can be spatially reused by multiple BSs who are sufficiently far away from each other.
The benefit of spectrum spatial reuse is characterized by the
frequency reuse factor $K$, which represents the proportion of the total
spectrum that one cell can utilize. For instance, $K=1$ means that any cell
can use all spectrum, and $K=1/7$ means that one cell can only utilize $1/7$ of the total spectrum,
to avoid excessive interference among adjacent cells.
A \emph{back-of-the-envelope} calculation suggests that, if the total number of required channels for all $N$  BSs is $C$,
then $\frac{C/N}{K}$ distinct radio channels are needed to serve the entire cellular
network.

In the case without D2D, the sum spectrum of all BSs is $F^{\textsf{ND}}$, which
corresponds to the total number of channels for all cells. Thus, with frequency reuse
factor $K$, $\frac{F^{\textsf{ND}}}{NK}$ distinct channels are needed without D2D.

In the case with D2D, D2D communication can degrade the original frequency reuse pattern
if they are sharing the same spectrum with cellular users (which is called underlay D2D \cite{Doppler09}).
Given the new frequency reuse factor $K^{\textsf{D2D}} (\le K)$. A back-of-the-envelope analysis suggests that
$\frac{F^{\textsf{D2D}}}{NK^{\textsf{D2D}}}$ distinct radio channels are needed with D2D load balancing. Consequently,
the spectrum reduction can be estimated as
\be
\frac{\frac{F^{\textsf{ND}}}{NK} - \frac{F^{\textsf{D2D}}}{NK^{\textsf{D2D}}}}{\frac{F^{\textsf{ND}}}{NK}}
= 1- \frac{K}{K^{\textsf{D2D}}} \times \frac{F^{\textsf{D2D}}}{F^{\textsf{ND}}}
= 1- \frac{K}{K^{\textsf{D2D}}} (1-\rho).
\label{equ:relation_spectrum_reduction}
\ee
Eq. \eqref{equ:relation_spectrum_reduction} suggests that our calculation of $\rho$ without frequency reuse
gives us a first-order understanding of how much spectrum reduction can be achieved by D2D load balancing with frequency reuse.

\begin{figure*}
  \centering
  \subfigure[Spectrum reduction v.s. $\lambda$.]{
    \label{fig:spectrum-reduction-vs-lambda} 
    \includegraphics[width=0.235\linewidth]{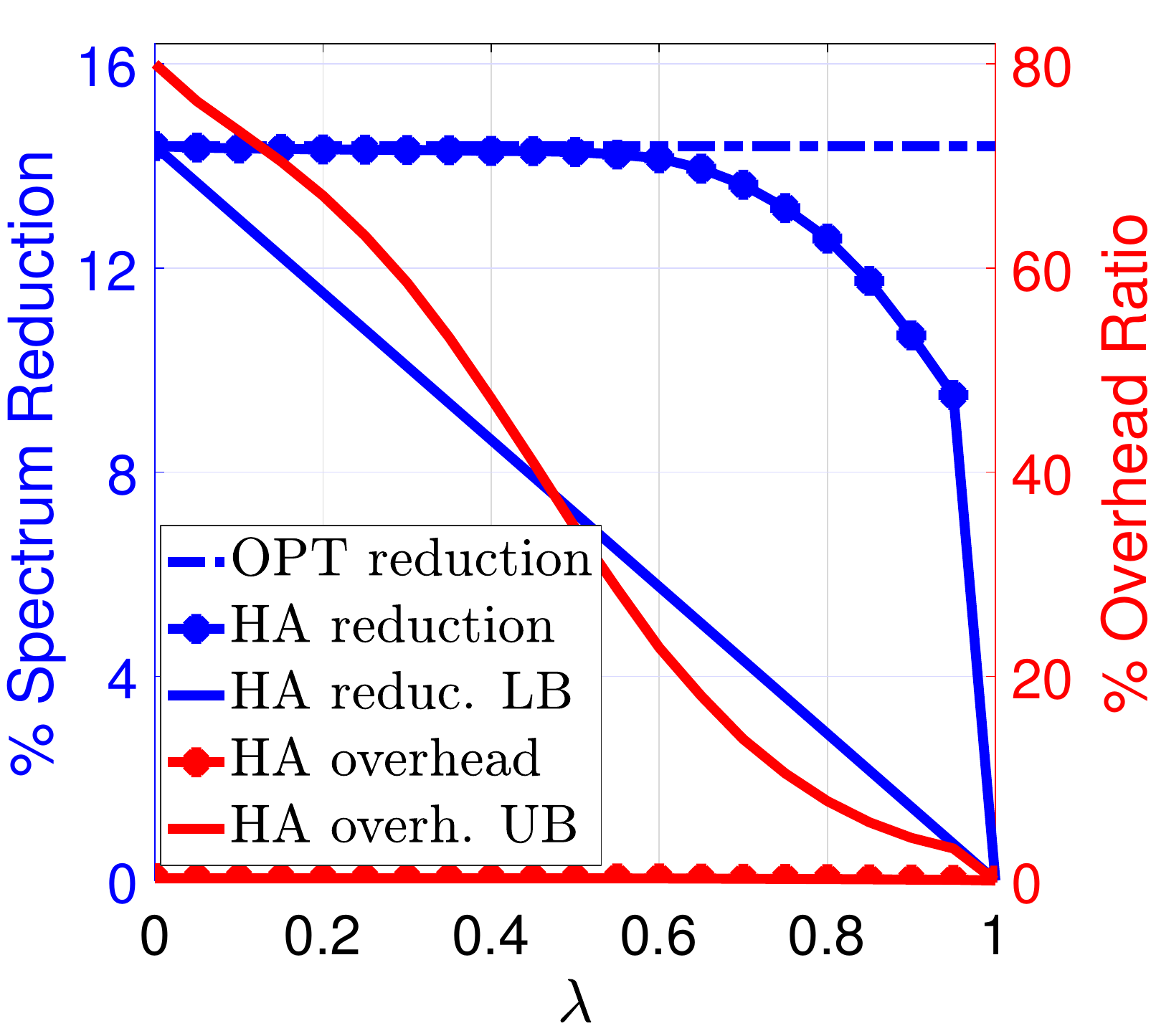}}
    \subfigure[Time/Space complexity v.s. $\lambda$.]{
    \label{fig:complexity-vs-lambda} 
    \includegraphics[width=0.235\linewidth]{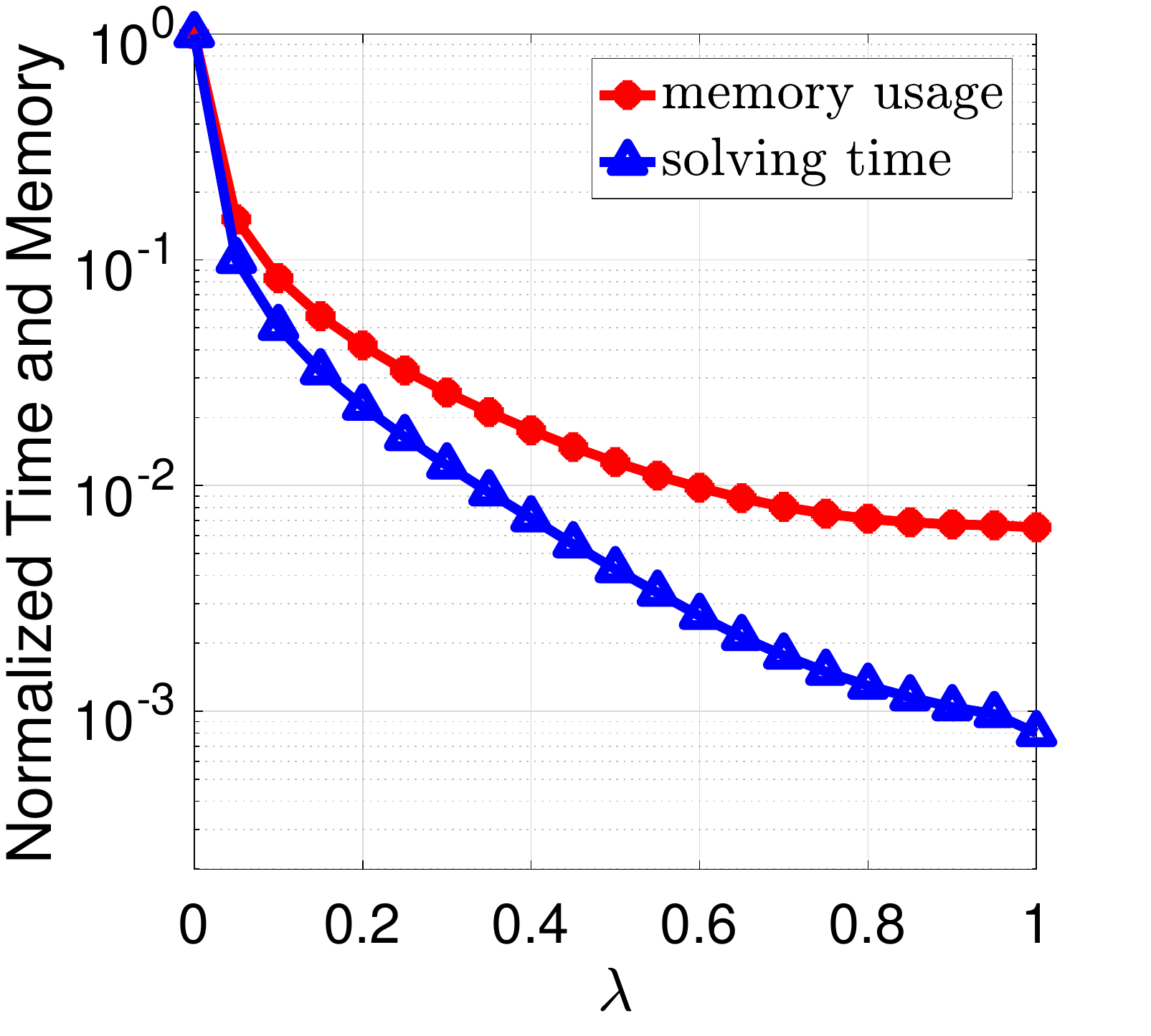}}
    \subfigure[Solving time.]{
    \label{fig:solving-time} 
    \includegraphics[width=0.235\linewidth]{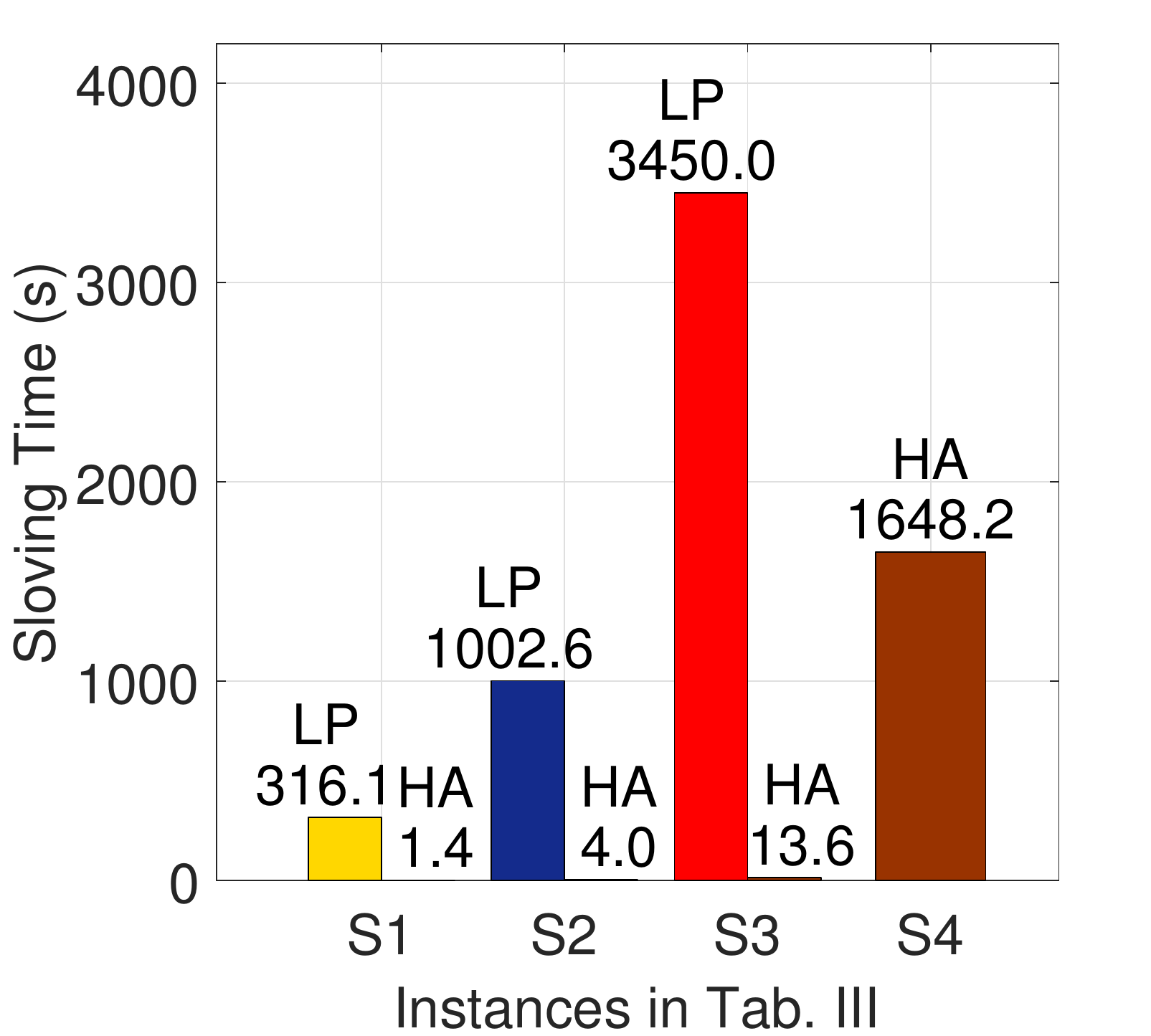}}
    \subfigure[Memory usage.]{
    \label{fig:memory-usage} 
    \includegraphics[width=0.235\linewidth]{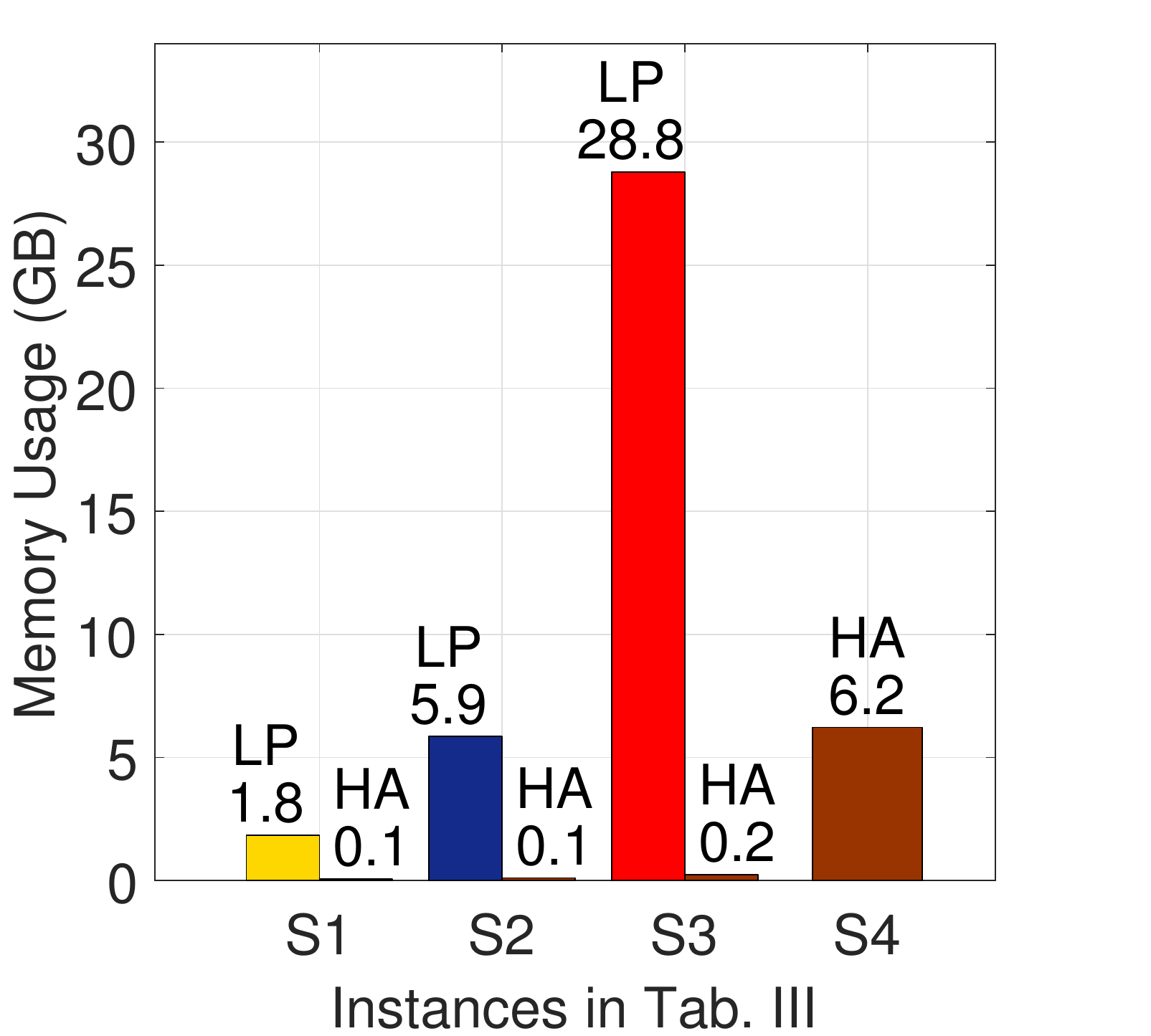}}
  \caption{Performance and complexity of our heuristic algorithm. Here,
  (a) and (b) show the performance and the complexity of the heuristic algorithm with different $\lambda$ values; (c) and (d) compare the solving time and memory usage of the global LP approach (LP)
and the heuristic algorithm (HA) with $\lambda=0.5$.}
  \label{fig:performance-complexity-heuristic} 
\end{figure*}

\section{Empirical Evaluations} \label{sec:simulation}
In this section, we use real-world 4G uplink traffic traces from Smartone, a major cellular network operator
in Hong Kong, to evaluate the performance of our proposed D2D load balancing scheme.

Our objectives are three-fold: (i) to evaluate the performance and complexity of our proposed low-complexity heuristic algorithm
in Sec.~\ref{sec:heuristic},
(ii) to evaluate the benefit in terms of spectrum reduction
and the cost in terms of D2D traffic overhead ratio of D2D load balancing scheme,
and (iii) to measure the impact of different system parameters.

\subsection{Methodology}
\textbf{Dataset:} Our Smartone dataset contains 510 cell sectors covering a highly-populated area of 22 km$^2$ in Hong Kong.
We merge them based on their unique site locations and get 152 BSs/cells. The data traffic traces
are sampled every 15 minutes, spanning a 29-day period from $2015/01/05$ to $2015/02/02$.

\textbf{Network Topology:} Each BS's location is its corresponding site location.
Each BS covers a circle area with radius 300m centered around its location. In each BS, 40 users are uniformly distributed
in the coverage circle. Assume that the communication range for all user-to-BS links
is 300m and the communication range for all D2D links is 30m. Then we can construct the cellular network topology
$\mathcal{G}=(\mathcal{V},\mathcal{E})$.
For each link $(u,v) \in \mathcal{E}$ with distance $d_{u,v}$, we use Shannon capacity to be the link rate, i.e.,
$R_{u,v} = \log_2(1 + {P_td_{u,v}^{-3.5}}/{N})$,
where $P_t=21$dBm is the transmit power and $N=-102$dBm is the noise power.

\textbf{Traffic Model:} We let each slot last for 2 seconds and
thus we have $T=24 \times 3600 /2 = 43200$ slots in each day.
Each data point in the raw traffic trace is the aggregate traffic volume of 15 minutes.
To get fine-granularity traffic demands, we randomly\footnote{When we say ``randomly", we draw a number from its range
uniformly.} generate 120 positive real numbers in $(0,1]$ and
then divide the aggregate traffic volume on a pro-rata basis according to the values of such 120 numbers.
Thus, we get 120 traffic demands of different volumes for each data point.
For each generated traffic demand $j$, we randomly assign it to a user $u_j$ from the total 40 users,
randomly set its start time $s_j$ from the total $15\times 60/2=450$ slots, and randomly
set its delay $(e_j-s_j+1)$ from the range $\{3,4,5\}$.

\textbf{Tools:} We use the state-of-the-art LP solver Gurobi \cite{gurobi} and
implement all evaluations with Python language.
All evaluations are running in a cluster of 30 computers, each of which
has a 8-core Intel Core-i7 3770 3.4Ghz CPU with 30GB memory, running
CentOS 6.4.

%
%
%

\begin{figure*}[t!]
\begin{minipage}[t]{0.3\linewidth}
  \centering
	\includegraphics[width=\linewidth]{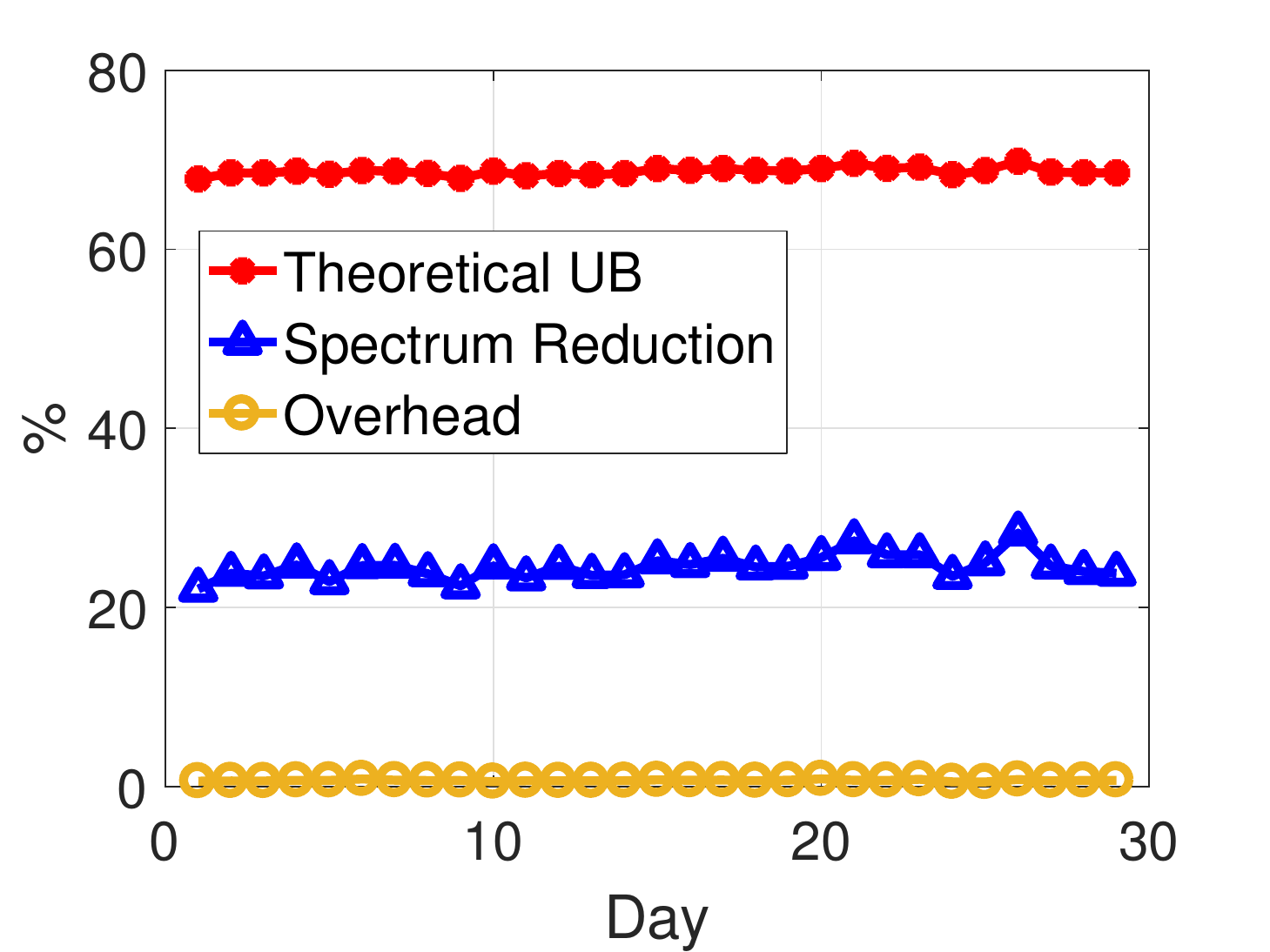}\\
	\caption{Spectrum reduction (ant its upper bound) and overhead ratio in 29 days.}\label{fig:spectrum-reduction_day}
\end{minipage}
\hfill
\begin{minipage}[t]{0.3\linewidth}
\centering
\includegraphics[width=\linewidth]{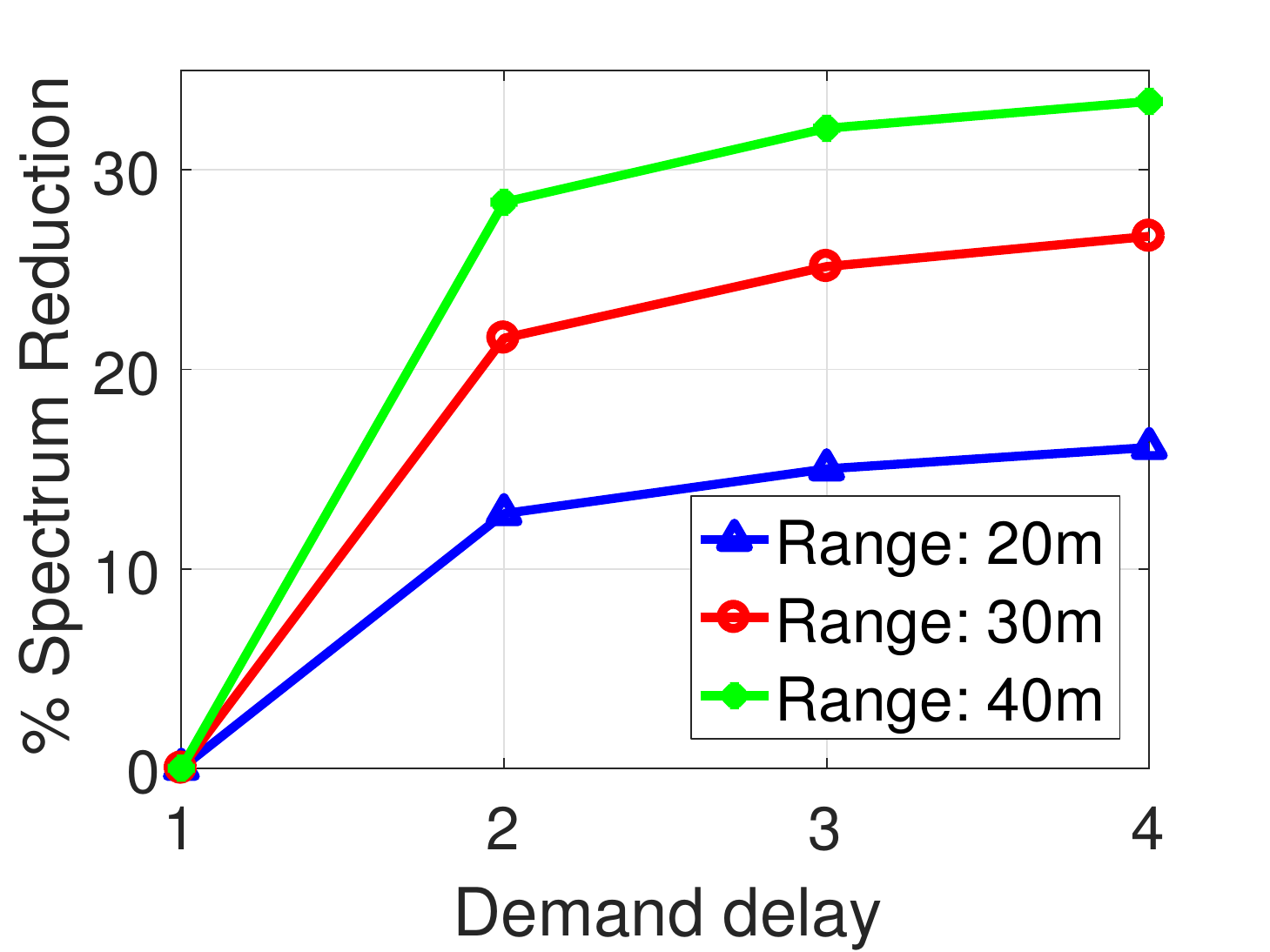}
\caption{\label{fig:impact-of-delay}  Impact of demand delay and D2D communication range.}
\end{minipage}
\hfill
\begin{minipage}[t]{0.3\linewidth}
\centering
\includegraphics[width=\linewidth]{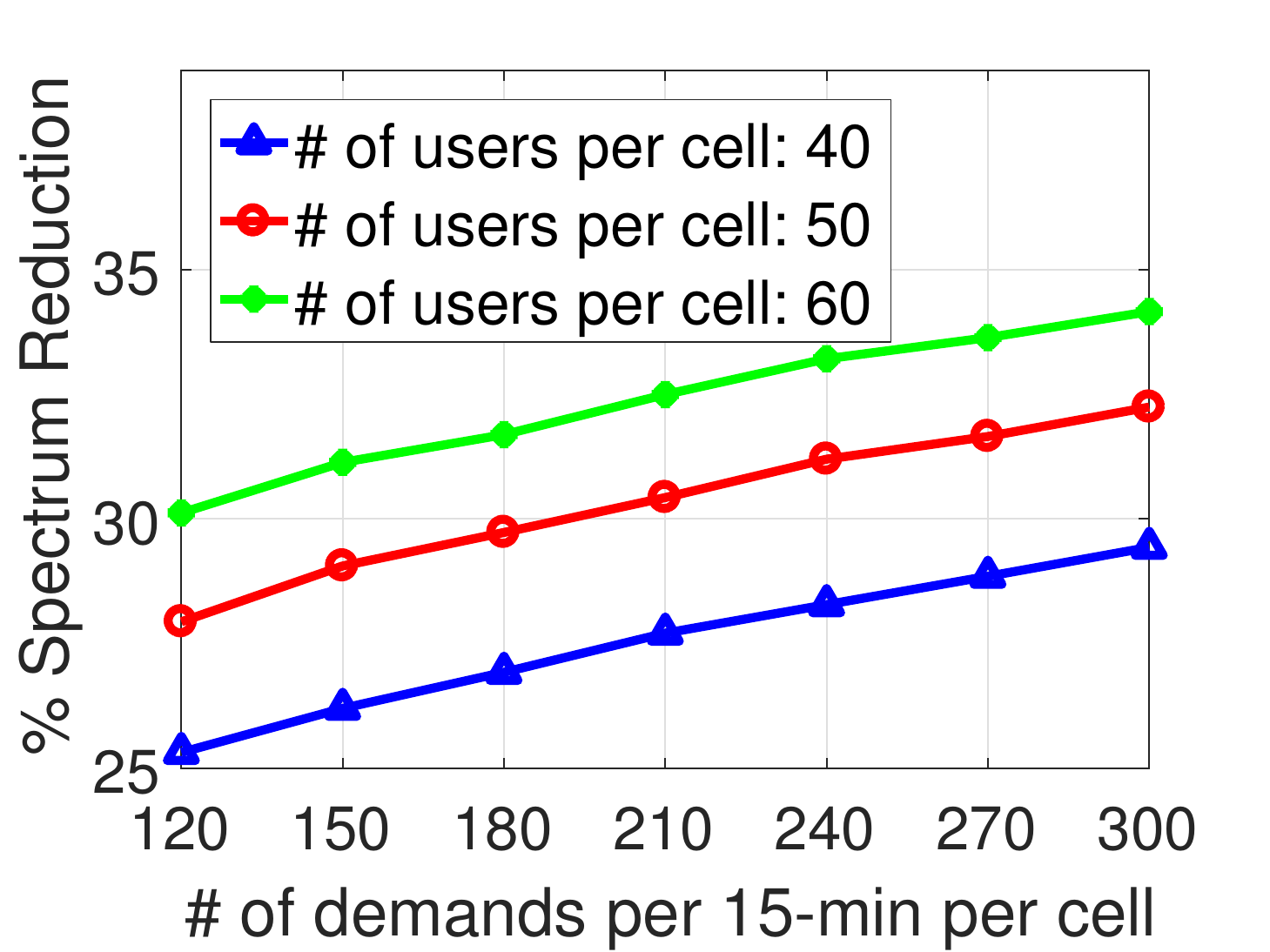}
\caption{\label{fig:impact-of-users}  Impact of user density and demand intensity.}
\end{minipage}
\end{figure*}

\subsection{Performance and Complexity of the Heuristic Algorithm}
As seen soon, our global LP approach cannot be applied to the whole cellular network due to its high complexity.
Instead, we should apply our low-complexity heuristic algorithm. In this section, we
show the performance and complexity of our heuristic algorithm and
hence justify why we can apply it to the whole cellular network.

The global LP approach is the benchmark to evaluate the heuristic algorithm but
we cannot use it for large-scale networks. We thus evaluate
them for small-scale networks. More specifically,
we divide the entire $22\text{km}^2$ region of 152 BSs into 22 small regions of 3 to 10 BSs.
For each small region and each day,
we use the global LP approach and the heuristic algorithm with different $\lambda$ values
to solve the problem $\textsf{Min-Spectrum-D2D}$ and get the spectrum reduction and the overhead ratio.
We then get the average spectrum reduction and average overhead ratio of both algorithms over
all 22 small regions and all 29 days, as shown in Fig. \ref{fig:spectrum-reduction-vs-lambda}.
Similarly, we show the normalized time/space complexity of our heuristic algorithm with different $\lambda$ values
in Fig.~\ref{fig:complexity-vs-lambda}

From Fig. \ref{fig:spectrum-reduction-vs-lambda} and Fig.~\ref{fig:complexity-vs-lambda},
we can see the tradeoff between performance (in terms of spectrum reduction) and the time/space complexity
controlled by parameter $\lambda$. Increasing $\lambda$ reduces the complexity but degrades the performance.
However, our heuristic algorithm achieves close-to-optimal performance when $\lambda$ is in $[0,0.5]$ and we can achieve
100x complexity reduction when we use $\lambda=0.5$.
Since our results in Fig. \ref{fig:spectrum-reduction-vs-lambda} and Fig.~\ref{fig:complexity-vs-lambda}
consider all 22 small regions of the entire region and all 29-day traffic traces,
it is reasonable to apply our heuristic algorithm
with $\lambda=0.5$ to the whole cellular network.
Thus, in the rest of this section, we set $\lambda=0.5$ for our heuristic algorithm

In Fig. \ref{fig:spectrum-reduction-vs-lambda}, we also show our spectrum reduction lower bound $(1-\lambda) \rho$ proposed in Theorem~\ref{thm:performance-heuristic}
and our overhead ratio upper bound $\frac{ (d_{\max}-1) \sum_{j \in \mathcal{J}^{\textsf{D2D}  }(\lambda)} r_j }{
(d_{\max}-1) \sum_{j \in \mathcal{J}^{\textsf{D2D}  }(\lambda)} r_j + \sum_{j \in \mathcal{J}} r_j}$ proposed in
Theorem~\ref{the:overhead-upper-bound-heuristic}. As we can see, we verify the correctness of both bounds.
More importantly, our empirical overhead ratio is much lower than the upper bound, almost close to 0, meaning
that we can achieve the spectrum reduction with very low overhead.

\begin{table}[t]
	\centering
	\caption{Four Different Problem Instances.}
	\label{tab:simulation_instances}
	\begin{tabular}{|l|l|l|l|l|l|l|l|}
		\hline
		Instance           & $|\mathcal{B}|$ & $|\mathcal{U}|$ & $|\mathcal{E}|$  & $|\mathcal{J}|$ & $\sum\limits_{b\in \mathcal{B}}|\mathcal{J}_b^{\text{D2D}}(\lambda)|$   & $T$     \\ \hline
		S1                 & 3    & 120  & 155   & 34080          & 182  & 43200 \\ \hline
		S2                 & 6    & 240  & 351   & 65520          &  377  & 43200 \\ \hline
		S3                 & 9    & 360  & 674   & 103680         & 632 & 43200 \\ \hline
		S4  			   & 152  & 6080 & 11794 & 1647480        & 11960  & 43200  \\ \hline
	\end{tabular}
\end{table}

To more concretely compare our heuristic algorithm (with $\lambda=0.5$) and our global LP approach,
we consider four different problem instances as shown in Tab. \ref{tab:simulation_instances}.
They have different number of BSs, users, links, and demands. Instance S4 is our whole cellular network.
We show their computational cost in Fig. \ref{fig:solving-time} and Fig. \ref{fig:memory-usage}.
From instances S1-S3, we can see that our heuristic algorithm has much lower time/space complexity than our global LP approach.
For our whole cellular network, i.e., instance S4, we cannot apply our global LP approach with our computational resources,
but our heuristic algorithm takes less than 30 minutes of time and consumes less than 6GB of memory.
The reason that we can get substantial complexity reduction is because the number of demands participating in
D2D load balancing in our heuristic algorithm, i.e.,  $\sum_{b\in\mathcal{B}}|\mathcal{J}^{\text{D2D}}_b(\lambda)|$,
is much smaller than the total number of demand, i.e., $|\mathcal{J}|$. As we can see from
Tab.~\ref{tab:simulation_instances}, $\sum_{b\in\mathcal{B}}|\mathcal{J}^{\text{D2D}}_b(\lambda)|$ is only about 0.7\% of $|\mathcal{J}|$
for instance S4.
%
%
%
%

\subsection{Spectrum Reduction and Overhead Ratio of D2D Load Balancing}

As justified in the previous subsection, we apply our heuristic algorithm with $\lambda=0.5$ to
the whole cellular network of all 152 BSs in the area of $22\text{km}^2$.
We show the 29-day spectrum reduction and overhead ratio in Fig. \ref{fig:spectrum-reduction_day}.
On average our proposed D2D load balancing scheme
can reduce spectrum by 25\% and the overhead ratio is only 0.7\%. Thus,
to serve the same set of traffic demands, cellular network operators like Smartone could
reduce its spectrum requirement by 25\% at the cost of negligible 0.7\% more D2D traffic
by using our D2D load balancing scheme.
Fig. \ref{fig:spectrum-reduction_day} also verifies the upper bound, represented in Theorem \ref{the:trivial_upper_bound} and Theorem \ref{the:ratio_bound}.
The average value of the upper bound of spectrum reduction is  68.69\%.

\subsection{Impact of System Parameters}
In this subsection, we evaluate the impact of four system parameters:
the demand delay, the D2D communication range, the number of users per cell (user density),
and the number of demands per cell per 15 minutes (demand intensity).
The results are shown in Fig.~\ref{fig:impact-of-delay} and Fig.~\ref{fig:impact-of-users}.
We observe that our D2D load balancing scheme brings more spectrum reduction
with larger demand delay, larger D2D communication range,
larger user density, or larger demand intensity.
The reason is as follows. Larger demand delay and larger demand intensity imply that traffic demands
can be balanced with more freedom, and larger D2D communication range and larger user density result in
better network connectivity, both of which enable D2D load balancing scheme to exploit more benefit.

\section{Conclusion and Future Work} \label{sec:conclusion}
To the best of our knowledge, this is the first work to characterize the system-level benefit and cost of D2D load balancing,
through both theoretical analysis and empirical evaluations. We show that
D2D load balancing can substantially reduce the spectrum requirement at low cost,
which provides strong support to standardize D2D in the coming cellular systems.
This work aims to provide performance metrics/benchmarks
and call for participation on the D2D load balancing scheme.
In the future, it is important and interesting to
jointly consider D2D load balancing and spectrum reuse/sharing among different cells and/or among different links,
design online and/or distributed traffic scheduling algorithms,
incorporate more realistic considerations such as transmission outage and user mobility,
and eventually implement the D2D load balancing scheme in practical systems.

\appendices


\ifx \ISTR \undefined
\else

\section{Case Study of Real-World 4G cellular data traffic traces} \label{app:case-study}
We carry out a case-study based on 4G
cell-traffic
traces from Smartone \cite{smartone} (this complements the study in our conference version~\cite{deng2015device}, which was based on 3G data traces),
a major cellular network operator
in Hong Kong, a highly-populated metropolis.
Smartone deploys 152 small-cell base stations in
the case-study area of 22 square kilometers, with cell radii of 200-300
meters. The traces include 4G data traffic
for each cell, sampled at 15-minute intervals over a
month in 2015. The results are shown in Fig.~\ref{fig:real_world_traffic}.

We have the following important observations.
\begin{itemize}
\item
First, the empirical CDF of the cell-capacity utilization
in Fig.~\ref{fig:low_temporal_utilization} shows that the average cell-capacity
utilization is 7.6\%,
and 90\% of the cells are less than 20\% utilized.
This confirms that small-cell architecture indeed causes very low
spectrum temporal utilization, and it suggests ample room to improve temporal
utilization.
\item
Second, from the 48-hour traffic plot of two adjacent cells in Fig.~\ref{fig:traffic_Imbalance},
we observe that their peak traffic occurs at different time epochs.
We remark that this observation is indeed common among the cells we studied.
We plot the CDF of Pearson correlation coefficients \cite{benesty2009pearson} of traffics
of all adjacent BS-pairs in Fig.~\ref{fig:low_temporal_correlation}. As we can see,
the average correlation is 9\% and more than 80\% of adjacent BS-pairs are less than 20\% correlated.
It implies that one may shift the peak traffic from a congested
cell to its under-utilized neighbors, so as to serve the traffic
without allocating extra spectrum, effectively improving the spectrum temporal utilization.
\end{itemize}

\begin{figure*}[t]
  \centering
    \subfigure[Empirical CDF for cell-capacity utilization of 152 cells for one month.]{
    \label{fig:low_temporal_utilization} 
    \includegraphics[width=0.3\linewidth]{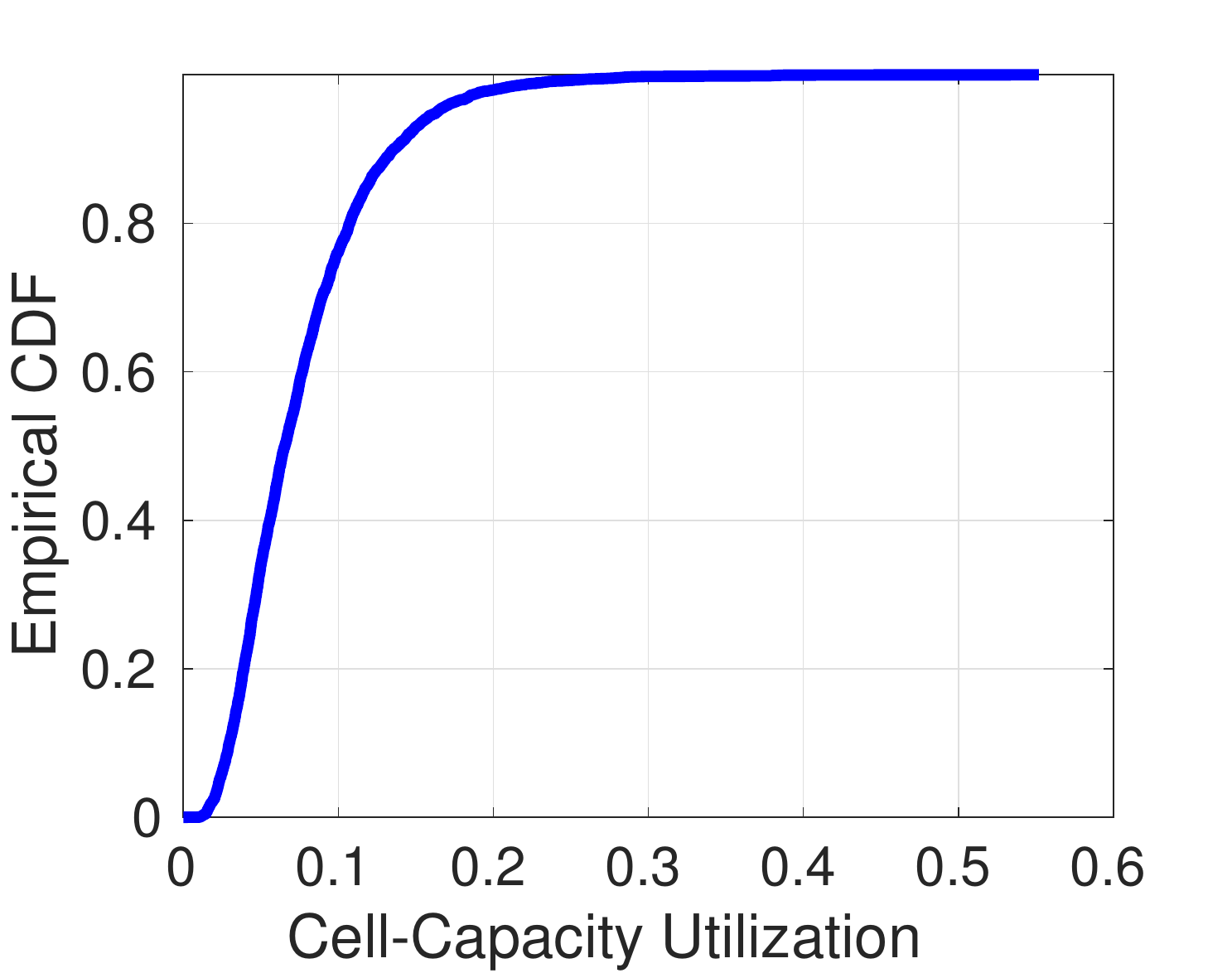}}
        \subfigure[Empirical CDF of Pearson correlation coefficients \cite{benesty2009pearson} of traffics of
adjacent BSs in one month.]{
    \label{fig:low_temporal_correlation} 
    \includegraphics[width=0.3\linewidth]{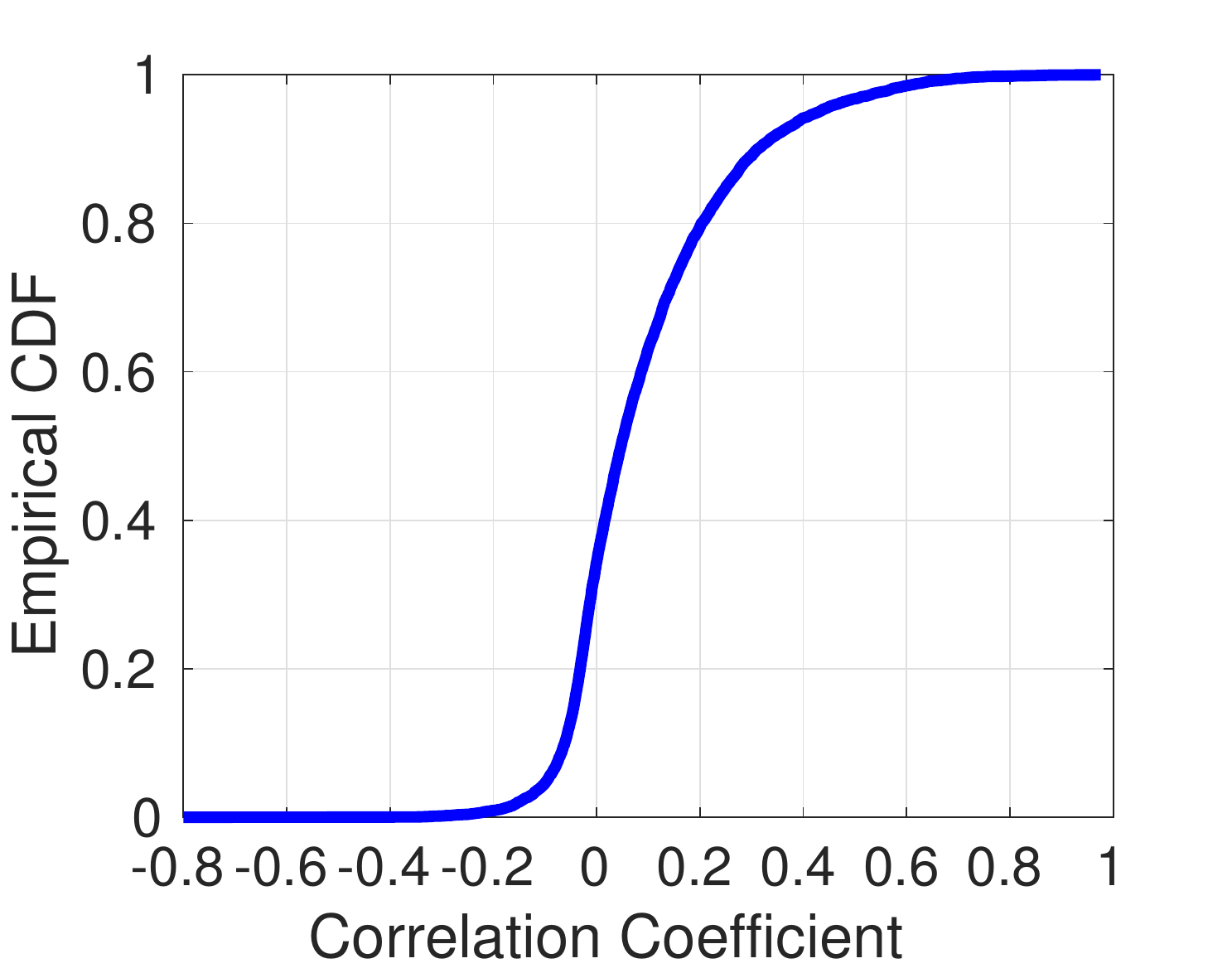}}
  \subfigure[4G (aggregated) mobile data traffic of two adjacent cells in 48 hours.]{
    \label{fig:traffic_Imbalance} 
    \includegraphics[width=0.3\linewidth]{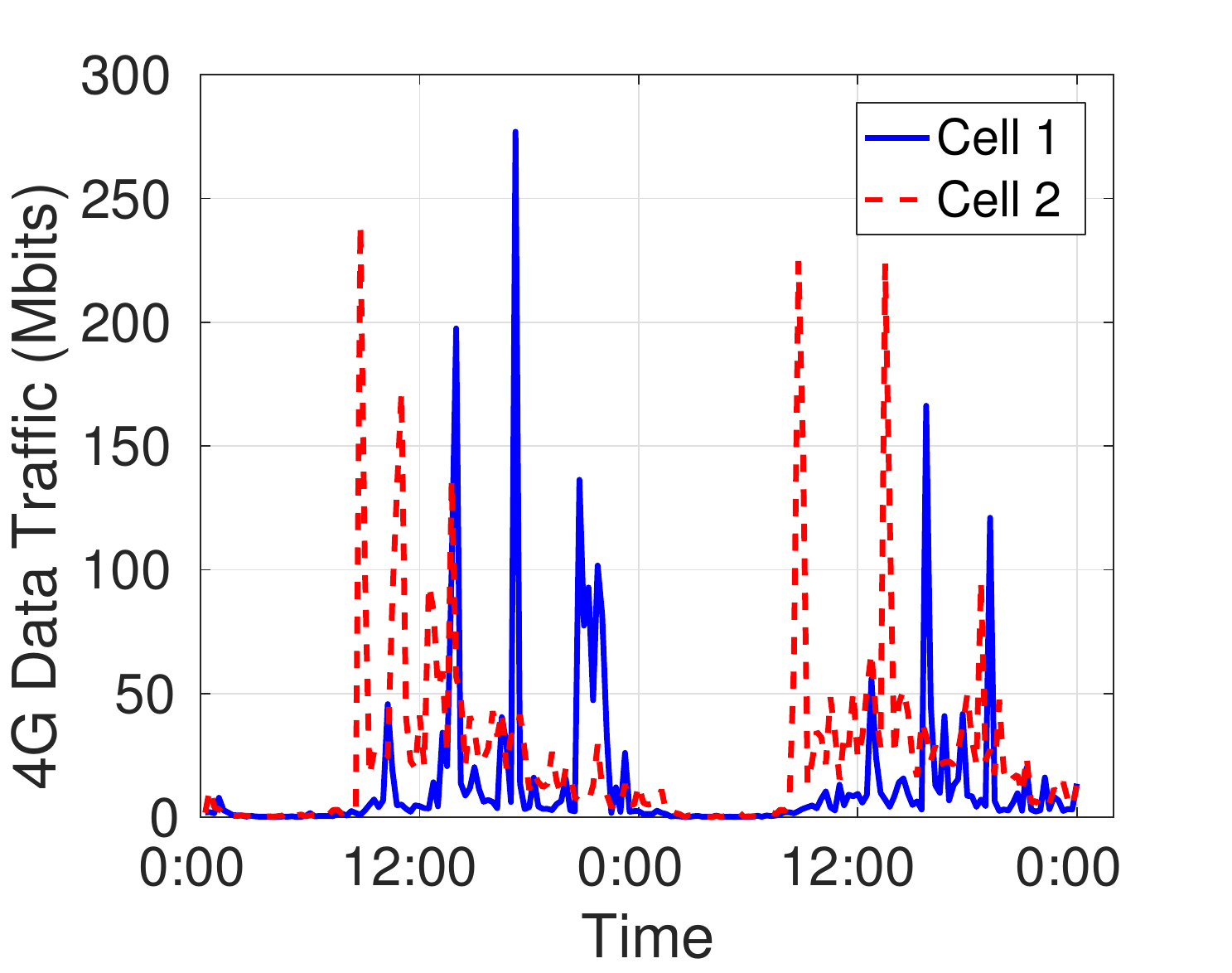}}
  \caption{Real-world 4G cellular data traffic traces.}
  \label{fig:real_world_traffic} 
\end{figure*}

\section{Reduce Complexity of \textsf{Min-Spectrum-D2D}} \label{sec:time_space_complexity}
To solve \textsf{Min-Spectrum-D2D} faster, we will use the following two implementation techniques
in space domain and time domain, respectively.
In space domain, we  reduce the memory usage by maintaining an available link list
for each traffic demand $j \in \mathcal{J}$. Since $j$ has a delay requirement of  $(e_j-s_j+1)$,
such traffic cannot reach too far away links.
Specifically, link $(u,v)$ is available for traffic demand $j$ only if the shortest
path of node $u_j$ and node $u$ is not larger than $(e_j-s_j)$. Therefore, we only need
to create the variable $x^{j}_{u,v}(t)$ for those available links $(u,v)$.

In the time domain, we can use multi-thread to speed up model-building time
when running in multi-processor operating system.
For the traffic scheduling policy constraints in \eqref{equ_demand}, \eqref{equ_reach},
\eqref{equ_conservation}, \eqref{equ:y_nonnegative}, and \eqref{equ:y_nonnegative_selflink}, different traffic demands
can run concurrently. For the peak traffic constraints in \eqref{equ:d2d_peak_cons1} and \eqref{equ:d2d_peak_cons2},
different BSs can run concurrently. Therefore, we can parallelize the constraint-building
process. Note that the Gurobi does not support multi-thread programming for a single environment.
One way to use multi-thread is to store a set of \texttt{GRBLinExpr} objects and return to the
main thread and pass them to the \texttt{GRBModel.addConstr()} function.

\section{Proof of {Theorem} \ref{the:YDS}} \label{app:proof_YDS}

Denote
\be
I^* = \operatorname*{arg\,max}_{I \subset [T]} g_b(I) = [z_1, z'_1].
\ee

First, we show that $F_b^{\textsf{ND}} \ge g_b(I^*)$. This is true because the feasible spectrum amount $F_b^{\textsf{ND}}$
can finish all traffic demands in the interval $I^*$, i.e., we must have
\be
(z'_1-z_1+1) F_b^{\textsf{ND}}  \ge \sum\limits_{j \in \mathcal{A}_b(I^*)} \frac{r_j}{R_{u_j,b}}.
\ee

Second, we show that $g_b(I^*)$ can finish all traffic in the interval $[T]$ with EDF, i.e.,
$F_b^{\textsf{ND}}  \le g_b(I^*)$.
This can be proved by contradiction. Suppose $g_b(I^*)$ cannot finish all traffic in the interval $[T]$. Then
we record the time when EDF returns false as $z_f$, which must be the deadline of a valid yet uncompleted traffic.
For any $t \in [z_f]$,
we define a binary variable $h_t$ to indicate whether or not the peak traffic is fully utilized as follows,
\be
h_t =
    \left\{
      \begin{array}{ll}
        1, & \hbox{if $\gamma_b(t)=g_b(I^*)$;}\\
        0, & \hbox{otherwise.}
      \end{array}
    \right.
\ee
Clearly we must have $h_{z_f} = 1$.
Now let us define $z_0$ as the latest time such that $h_t = 0$, i.e.,
$z_0 = \max\limits_{t \in [z_f]: h_t = 0} t$. If $h_t = 1$
for any $t \in [z_f]$, then we let $z_0 = 0$. Since $h_{z_0} = 0$,
we conclude that all traffic demands whose deadlines are not
larger than $z_0$ have been completed at the end of slot $z_0$ with EDF algorithm.
Then we consider the interval $I' = [z_0+1, z_f]$.
Since $h_t = 1 $ for any $t \in [z_0+1, z_f]$, we obtain that the total traffic volume
delivered in the interval $I'$ is
$(z_f - z_0)g(I^*)$. Since EDF returns false at the end of slot $z_f$, we must have
\be
(z_f - z_0)g_b(I^*) <  \sum\limits_{j \in \mathcal{A}_b(I')} \frac{r_{j}}{R_{u_j,b}},
\ee
which yields to
\be
g_b(I') = \frac{\sum\limits_{j \in \mathcal{A}_b(I')} \frac{r_j{j}}{R_{u_j,b}}}{z_f - z_0} > g_b(I^*).
\ee
This is a contradiction to the fact that $I^*$ maximize $g_b(I)$.

Therefore, $F^{\textsf{ND}}_b = g_b(I^*)$.

\section{Proof of Theorem \ref{the:trivial_upper_bound}} \label{app:proof_trivial_upper_bound}
Let us denote the original problem instance by $P$, whose minimum spectrum to serve all
traffic with D2D load balancing is $F^{\textsf{D2D}}$. Now we construct a new problem instance $P'$,
which has the same network topology as the original problem instance $P$.
However, $P'$ differs from $P$ in the following three aspects:
\begin{enumerate}[(i)]
\item the link rate of any user-to-BS link is set as $R_{\max}$, which is larger than (or at least equal to)
that in $P$;
\item the link rate of any D2D link is set as $+\infty$, implying that D2D communication does not consume any spectrum resources;
\item any D2D transmission does not incur any delay.
\end{enumerate}

Clearly, the minimum spectrum to serve all traffic demands with D2D in $P'$, denoted by $F'$,
is less than that in $P$, i.e.,
\be
F' \le F^{\textsf{D2D}}.
\label{equ:F-inequ}
\ee

Now we further construct another problem instance $P''$ as follows:
\begin{enumerate}[(i)]
\item It has only one (grand) BS $b_0$
\item It has all users $\mathcal{U}$ in the original problem instance $P$
\item All users connect to the grand BS $b_0$ with link rate $R_{\max}$.
\item There are no D2D links.
\end{enumerate}

We denote the minimum spectrum to serve all traffic demands in $P''$ as $\underline{F}^{\textsf{D2D}}$.
Since $P''$ is just the single-BS case without D2D as studied in Sec.~\ref{sec:optimal_no_d2d},
we have $\underline{F}^{\textsf{D2D}}=\max_{I \subset [T]} g(I)$ where $g(I)$ is defined in \eqref{equ:intensity_d2d}.

In $P'$, any traffic volume traveling through one or multiple D2D links before reaching a BS (say BS $b$) will
only consume spectrum resources and incur delay in the last user-to-BS link; it is as if
we directly transmit such traffic volume to BS $b$. Therefore, problem instance $P'$ has the same minimum
spectrum as problem instance $P''$, i.e., $F'=\underline{F}^{\textsf{D2D}} =\max_{I \subset [T]} g(I)$.
Thus, from \eqref{equ:F-inequ}, we have
\be
\rho = \frac{F^{\textsf{ND}}-F^{\textsf{D2D}}}{F^{\textsf{ND}}} \le   \frac{F^{\textsf{ND}}-F'}{F^{\textsf{ND}}} = \frac{F^{\textsf{ND}}-\underline{F}^{\textsf{D2D}}}{F^{\textsf{ND}}}.
\ee

\fi

\section{Proof of {Theorem} \ref{the:ratio_bound}} \label{app:proof_upper_bound}



The proof logic is to construct a feasible solution to $\textsf{Min-Spectrum-ND}_b$
based on the optimal solution with D2D.
Let us denote the optimal traffic scheduling policy for $\textsf{Min-Spectrum-D2D}$
as $x_{u,v}^{j}(t)$ and the optimal spectrum amount for each BS $b$ as $F^{\textsf{D2D}}_b$.
Then consider BS $b \in \mathcal{B}$. For each traffic
demand $j \in \mathcal{J}_b$, user $s \in \mathcal{U}_b$ must transmit all volume $r_j$ either to BS $b$ directly
and/or to any other neighbour users via D2D links. Thus $\forall j \in \mathcal{J}_b$,
the following equality holds,
\bee
& r_j =  \sum_{t=s_j}^{e_j}  [ x_{u_j,b}^{j}(t)R_{u_j,b}  + \sum_{v: v \in \mathcal{U}_b, (u_j,v) \in \mathcal{E}} x^{j}_{u_j,v}(t)R_{u_j,v} \nnb \\
& + \sum_{b': (b,b') \in \mathcal{E}^{\textsf{D2D}}} \sum_{v: v \in \mathcal{U}_{b'}, (u_j,v) \in \mathcal{E}} x^{j}_{u_j,v}(t)R_{u_j,v}], \nnb
\eee
In addition, the (peak) spectrum requirement should be satisfied,
\be
\sum_{j \in \mathcal{J}_b} x^{j}_{u_j,b}(t)
+
\sum_{u \in \mathcal{U}_b} \sum_{v \in \text{in}\left( u \right)
\backslash \left\{ u \right\}} \sum_{j\in \mathcal{J}: t \in [s_j,e_j]} x^{j}_{v,u}(t)
\le F_b^{\textsf{D2D}},
\nnb
\ee
Now we construct a feasible solution to $\textsf{Min-Spectrum-ND}_b$, i.e., for any $j \in \mathcal{J}_b$,
\bee
& \bar{x}_{u_j,b}^{j}(t) =    x_{u_j,b}^{t}(t)  + \sum_{v: v \in \mathcal{U}_b, (u_j,v) \in \mathcal{E}} x^{j}_{u_j,v}(t)\frac{R_{u_j,v}}{R_{u_j,b}} \nnb \\
& + \sum_{b': (b,b') \in \mathcal{E}^{\textsf{D2D}}}
\sum_{v: v \in \mathcal{U}_{b'}, (u_j,v) \in \mathcal{E}} x^{j}_{u_j,v}(t) \frac{R_{u_j,v}}{R_{u_j,b}},
\eee

Thus we have
\bee
 & \gamma_b(t)= \sum_{j \in \mathcal{J}_b: t \in [s_j,e_j] } \bar{x}^{j}_{u_j,b}(t)
 = \sum_{j \in \mathcal{J}_b: t \in [s_j,e_j] }
[ x_{u_j,b}^{t}(t)  \nnb \\
& \qquad  + \sum_{v: v \in \mathcal{U}_b, (u_j,v) \in \mathcal{E}} x^{j}_{u_j,v}(t)\frac{R_{u_j,v}}{R_{u_j,b}} \nnb \\
& \qquad + \sum_{b': (b,b') \in \mathcal{E}^{\textsf{D2D}}}
\sum_{v: v \in \mathcal{U}_{b'}, (u_j,v) \in \mathcal{E}} x^{j}_{u_j,v}(t) \frac{R_{u_j,v}}{R_{u_j,b}}] \nnb \\
& \le \sum_{j \in \mathcal{J}_b: t \in [s_j,e_j] }
[ x_{u_j,b}^{t}(t) + r_{u_j} \sum_{v: v \in \mathcal{U}_b, (u_j,v) \in \mathcal{E}} x^{j}_{u_j,v}(t) \nnb \\
& \qquad + \sum_{b': (b,b') \in \mathcal{E}^{\textsf{D2D}}}
\tilde{r}^{b'}_{u_j} \sum_{v: v \in \mathcal{U}_{b'}, (u_j,v) \in \mathcal{E}} x^{j}_{u_j,v}(t)  ] \nnb \\
& \overset{(a)}{\le} \max\{r,1\} \sum_{j \in \mathcal{J}_b: t \in [s_j,e_j] }
[ x_{u_j,b}^{t}(t) +  \sum_{v: v \in \mathcal{U}_b, (u_j,v) \in \mathcal{E}} x^{j}_{u_j,v}(t)] \nnb \\
& \qquad + \tilde{r} \sum_{b': (b,b') \in \mathcal{E}^{\textsf{D2D}}} \sum_{j \in \mathcal{J}_b: t \in [s_j,e_j] }
 \sum_{v: v \in \mathcal{U}_{b'}, (u_j,v) \in \mathcal{E}} x^{j}_{u_j,v}(t) \nnb \\
& \le \max\{r,1\} F_b^{\textsf{D2D}} + \tilde{r} \sum_{b': (b,b') \in \mathcal{E}^{\textsf{D2D}}} F^{\textsf{D2D}}_{b'},
\eee
where $(a)$ trivially holds for $r > 1$ and also holds for $r \le 1$ by noting that there is no intra-cell
D2D traffic when $r \le 1$.
Therefore, $\max\{r,1\} F_b^{\textsf{D2D}} + \tilde{r} \sum_{b': (b,b') \in \mathcal{E}^{\textsf{D2D}}} F^{\textsf{D2D}}_{b'}$
is a feasible spectrum amount for BS $b$ without D2D. Thus
we must have
\be
F_b^{\textsf{ND}} \le \max\{r,1\} F_b^{\textsf{D2D}} + \tilde{r} \sum_{b': (b,b') \in \mathcal{E}^{\textsf{D2D}}} F^{\textsf{D2D}}_{b'}.
\ee

Then we do summation over all BSs and get
\bee
& F^{\textsf{ND}}  = \sum_{b \in \mathcal{B}} F^{\textsf{ND}}_b \nnb \\
& \le \sum_{b \in \mathcal{B}} \max\{r,1\} F_b^{\textsf{D2D}}
+ \tilde{r} \sum_{b \in \mathcal{B}} \sum_{b': (b,b') \in \mathcal{E}^{\textsf{D2D}}} F^{\textsf{D2D}}_{b'} \nnb \\
& \overset{(b)}{=} \max\{r,1\} \sum_{b \in \mathcal{B}} F_b^{\textsf{D2D}}
+ \tilde{r} \sum_{b' \in \mathcal{B}} \sum_{b: (b,b') \in \mathcal{E}^{\textsf{D2D}}} F^{\textsf{D2D}}_{b'} \nnb \\
& = \max\{r,1\} \sum_{b \in \mathcal{B}} F_b^{\textsf{D2D}}
+ \tilde{r} \sum_{b' \in \mathcal{B}} \delta_{b'}^- F_{b'}^{D2D} \nnb \\
& \le \max\{r,1\} \sum_{b \in \mathcal{B}} F_b^{\textsf{D2D}} + \tilde{r} \sum_{b' \in \mathcal{B}} \Delta^- F_{b'}^{\textsf{D2D}} \nnb \\
& = [\max\{r,1\} + \tilde{r} \Delta^-] F^{\textsf{D2D}},
\eee
where $(b)$ holds because any $(b,b') \in \mathcal{E}^{\textsf{D2D}}$ contributes one $\tilde{r}F_{b'}^{\textsf{D2D}}$ on both sides.
Thus, we conclude that
\be
\rho = \frac{F^{\textsf{ND}}-F^{\textsf{D2D}}}{F^{\textsf{ND}}}
\le \frac{\max\{r,1\} + \tilde{r} \Delta^--1}{\max\{r,1\} + \tilde{r} \Delta^-}.
\ee

\ifx \ISTR \undefined
\else
\section{Proof of {Fact \ref{fact:bs_ring}}} \label{app:proof_ring}
In the ring topology, we assume the BS is indexed from 1 to $N=2D-1$ counterclockwise.
In the case without D2D load balancing,
the minimum peak traffic for any BS $i \in [N]$  is
\be
F^{\textsf{ND}}_i = \frac{V}{D} \triangleq F^{\textsf{nd}}.
\ee

In the case with D2D load balancing, we will construct a traffic scheduling policy to achieve
the (peak) spectrum reqirement for any BS $i \in [N]$,
\be
F^{\textsf{D2D}}_i = \frac{V}{3D-2} \triangleq F^{\textsf{d2d}}.
\ee

Let us consider BS 1 firstly. For the traffic in BS 1,  we first consider the counterclockwise side,
i.e., $ 1 \to 2 \to 3 \to \cdots \to D $. We construct the following traffic scheduling policy
from slot 1 to slot $D$ where $b_i$ means BS $i$ and the $t$-th entry in the braces is the traffic volume at slot $t$ on that link:
\begin{itemize}
\item
$u_1 \to u_2: \quad \{\underbrace{F^{\textsf{d2d}}, \cdots, F^{\textsf{d2d}}}_{D-1}, 0\}$,
$u_2 \to b_2: \quad \{\underbrace{0, \cdots, 0}_{D-1},F^{\textsf{d2d}}\}$,
\item
$u_2 \to u_3: \quad \{0, \underbrace{F^{\textsf{d2d}}, \cdots, F^{\textsf{d2d}}}_{D-2}, 0\}$,
$u_3 \to b_3: \quad \{\underbrace{0, \cdots, 0}_{D-1},F^{\textsf{d2d}}\}$,
\item
$\cdots\cdots$
\item
$u_{D-1} \to u_D: \quad \{\underbrace{0,\cdots,0}_{D-2}, F^{\textsf{d2d}}, 0\}$,
$u_D \to b_D: \quad \{\underbrace{0, \cdots, 0}_{D-1},F^{\textsf{d2d}}\}$.
\end{itemize}

Clearly, the  counterclockwise side BSs can help transfer $(D-1)F^{\textsf{d2d}}$ traffic for user $u_1$.
We can construct the same traffic scheduling for the clockwise side, i.e., $1 \to (2D-1) \to (2D-2) \to \cdots \to (D+1)$
such that they also help transfer $(D-1)F^{\textsf{d2d}}$ traffic for user $u_1$.
In addition, user $u_1$ can directly transmit $DF^{\textsf{d2d}}$  traffic to BS 1 as
\begin{itemize}
\item
$u_1 \to b_1: \quad \{\underbrace{F^{\textsf{d2d}}, F^{\textsf{d2d}}, \cdots, F^{\textsf{d2d}}}_{D}\}$.
\end{itemize}

Hence, all the traffic for user $u_1$ has been finished before its deadline (slot $D$) because
\bee
DF^{\textsf{d2d}} + (D-1)F^{\textsf{d2d}} + (D-1)F^{\textsf{d2d}} = (3D-2)F^{\textsf{d2d}}
= V.
\nnb
\eee
Furthermore, we can check that the (peak) spectrum requirement for all $N$ BSs is $F^{\textsf{d2d}} = \frac{V}{3D-2}$.

In addition, since the ring topology is symmetric and all traffic is decoupled,
we immediately get that all other traffic can be satisfied when the spectrum amount for all BSs is $F^{\textsf{d2d}}$.

Therefore, we get the spectrum reduction
\bee
\rho  = \frac{F^{\textsf{ND}}-F^{\textsf{D2D}}}{F^{\textsf{ND}}} = \frac{NF^{\textsf{nd}}-NF^{\textsf{d2d}}}{NF^{\textsf{nd}}}
=  \frac{2(D-1)}{3D-2} \to \frac{2}{3} \; (D \to \infty).
\nnb
\eee

In addition, the sum D2D traffic for all users is,
\bee
V^{\textsf{D2D}} & =  N \cdot 2(F^{\textsf{D2D}} + 2F^{\textsf{d2d}} + \cdots + (D-1) F^{\textsf{d2d}}) \nnb \\
& = 2NF^{\textsf{d2d}} \sum_{i=1}^{D-1} i =  (D-1)D \cdot \frac{NV}{3D-2},
\nnb
\eee
and the sum traffic directly sent by users to BSs is the total traffic volume for all users
in the given traffic demand pattern, i.e.,
$V^{\textsf{BS}} = NV.$
Thus, the overhead ratio is
\bee
\eta  = \frac{V^{\textsf{D2D}}}{V^{\textsf{D2D}}+V^{\textsf{BS}}}
 = \frac{D(D-1)}{D^2+2D-2}.
 \nnb
\eee
The proof is completed.

\section{Proof of {Fact \ref{fact:bs_complete}}} \label{app:proof_complete}
In the case without D2D load balancing, the minimum (peak) spectrum requirement for any BS $i \in [N]$ is
\be
F^{\textsf{ND}}_i = \frac{V}{D} \triangleq F^{\textsf{nd}}.
\ee

In the case with D2D load balancing, we will construct a traffic scheduling policy to achieve
the (peak) spectrum requirement for any BS $i \in [N]$,
\be
F^{\textsf{D2D}}_i = \frac{2V}{(N+1)D} \triangleq F^{\textsf{d2d}}.
\ee

We first consider the traffic for user $u_1$ and construct the following
traffic scheduling policy:
\begin{itemize}
\item Case 1 when $D$ is even:
$\forall i \in [2,N]$, \\
$u_1 \to u_i: \quad \{\underbrace{F^{\textsf{d2d}}, \cdots, F^{\textsf{d2d}}}_{D/2}, \underbrace{0, \cdots, 0}_{D/2}\}$, \\
$u_i \to b_i: \quad \{\underbrace{0, \cdots, 0}_{D/2}, \underbrace{F^{\textsf{d2d}}, \cdots, F^{\textsf{d2d}}}_{D/2}\}$.
\item Case 2 when $D$ is odd:
$\forall i \in [2,N]$, \\
$u_1 \to u_i: \quad \{\underbrace{F^{\textsf{d2d}}, \cdots, F^{\textsf{d2d}}}_{(D-1)/2}, \frac{F^{\textsf{d2d}}}{2}, \underbrace{0, \cdots, 0}_{(D-1)/2}\}$, \\
$u_i \to b_i: \quad \{\underbrace{0, \cdots, 0}_{(D-1)/2}, \frac{F^{\textsf{d2d}}}{2}, \underbrace{F^{\textsf{d2d}}, \cdots, F^{\textsf{d2d}}}_{(D-1)/2}\}$.
\end{itemize}
In both cases, any other BS $i \in [2,N]$ can help transfer $\frac{D}{2} F^{\textsf{d2d}}$ traffic for user $u_1$.
Besides, user $u_1$ can transmit $DF^{\textsf{d2d}}$ traffic to BS 1 as:
\begin{itemize}
\item
$u_1 \to b_1: \quad \{\underbrace{F^{\textsf{d2d}}, \cdots, F^{\textsf{d2d}}}_{D}\}$.
\end{itemize}
Then we can check all traffic for user $u_1$
has been finished before the deadline (slot $D$) because
\bee
 DF^{\textsf{d2d}} + (N-1) \frac{D}{2} F^{\textsf{d2d}} = \frac{N+1}{2}DF^{\textsf{d2d}}
= V.
\nnb
\eee

In addition, we can see that the (peak) spectrum requirement for all BSs is $F^{\textsf{d2d}}$.

Since the complete topology is symmetric and all traffic is decoupled,
the traffic for all other users can be satisfied when the spectrum amount  for all BSs is $F^{\textsf{d2d}}$.

Therefore, the sum spectrum reduction is
\bee
\rho  = \frac{F^{\textsf{ND}}-F^{\textsf{D2D}}}{F^{\textsf{ND}}} = \frac{NF^{\textsf{nd}} - NF^{\textsf{d2d}}}{NF^{\textsf{nd}}}
= \frac{N-1}{N+1}\to 1 \; (N \to \infty).
\nnb
\eee

In addition, the sum D2D traffic for all users is,
\be
V^{\textsf{D2D}} = N \cdot (N-1)  \frac{D}{2}  F^{\textsf{d2d}} = \frac{N(N-1)V}{N+1},
\nnb
\ee
and the sum traffic directly sent by users to BSs is the total traffic volume for all users
in the given traffic demand pattern, i.e.,
$
V^{\textsf{BS}} = NV.
$
Thus, the overhead ratio is,
\be
\eta = \frac{V^{\textsf{D2D}}}{V^{\textsf{D2D}} + V^{\textsf{BS}}} = \frac{ \frac{N(N-1)V}{N+1}}{ \frac{N(N-1)V}{N+1} + NV} = \frac{N-1}{2N}.
\ee
The proof is completed.

\section{Proof of Theorem~\ref{the:overhead-upper-bound}} \label{app:proof-of-overhead-upper-bound}
First of all, each bit of traffic demand $j$ can at most travel $e_j-s_j$ times over D2D links before it
reaches a BS. Thus, each traffic demand $j$ can at most incur D2D traffic $r_j(e_j-s_j) \le r_j (d_{\max}-1)$.
The total D2D traffic is thus upper bounded by
\be
V^{\textsf{D2D}} \le (d_{\max}-1) \sum_{j \in \mathcal{J}} r_j = (d_{\max}-1) V^{\textsf{BS}}.
\ee
And thus the overhead ratio is upper bounded by
\be
\eta = \frac{V^{\textsf{D2D}}}{V^{\textsf{D2D}} + V^{\textsf{BS}}} \le \frac{d_{\max}-1}{d_{\max}}.
\ee

\section{An example for Our Heuristic Algorithm in Sec.~\ref{sec:heuristic}} \label{app:an-example-for-heuristic}

\begin{figure*}
  \centering
  \subfigure[Step I]{
    \label{fig:heuristic_Part1} 
    \includegraphics[width=0.3\linewidth]{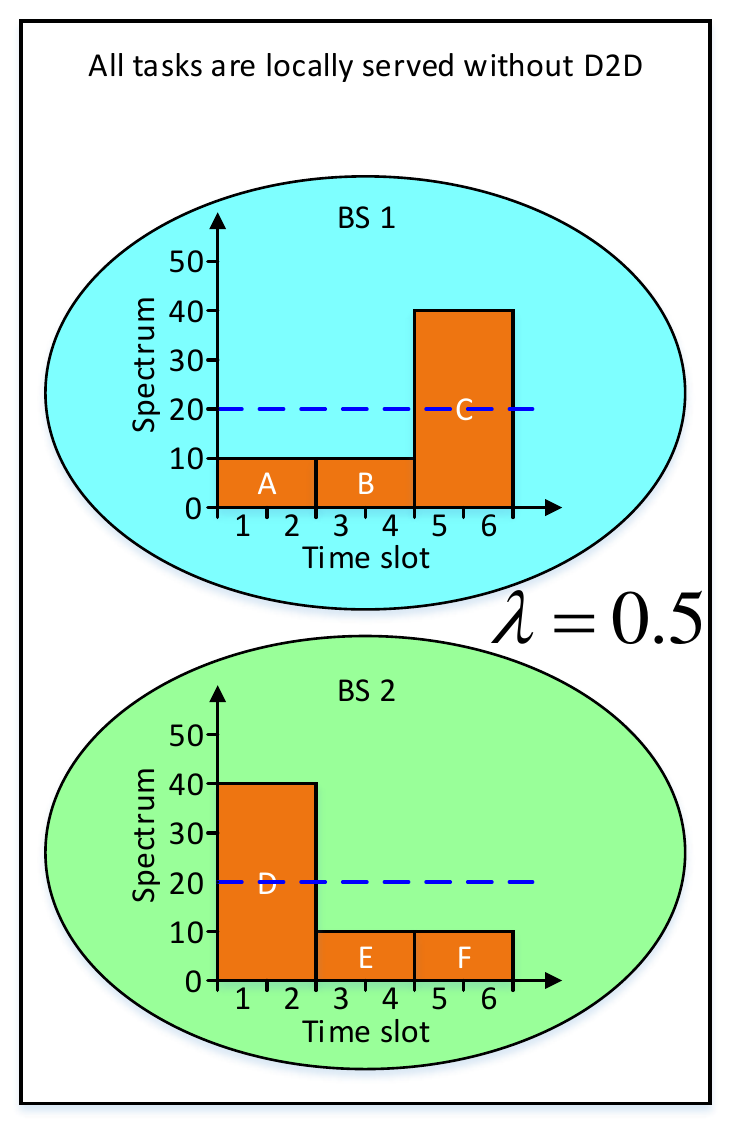}}
  \subfigure[Step II]{
    \label{fig:heuristic_Part2} 
    \includegraphics[width=0.3\linewidth]{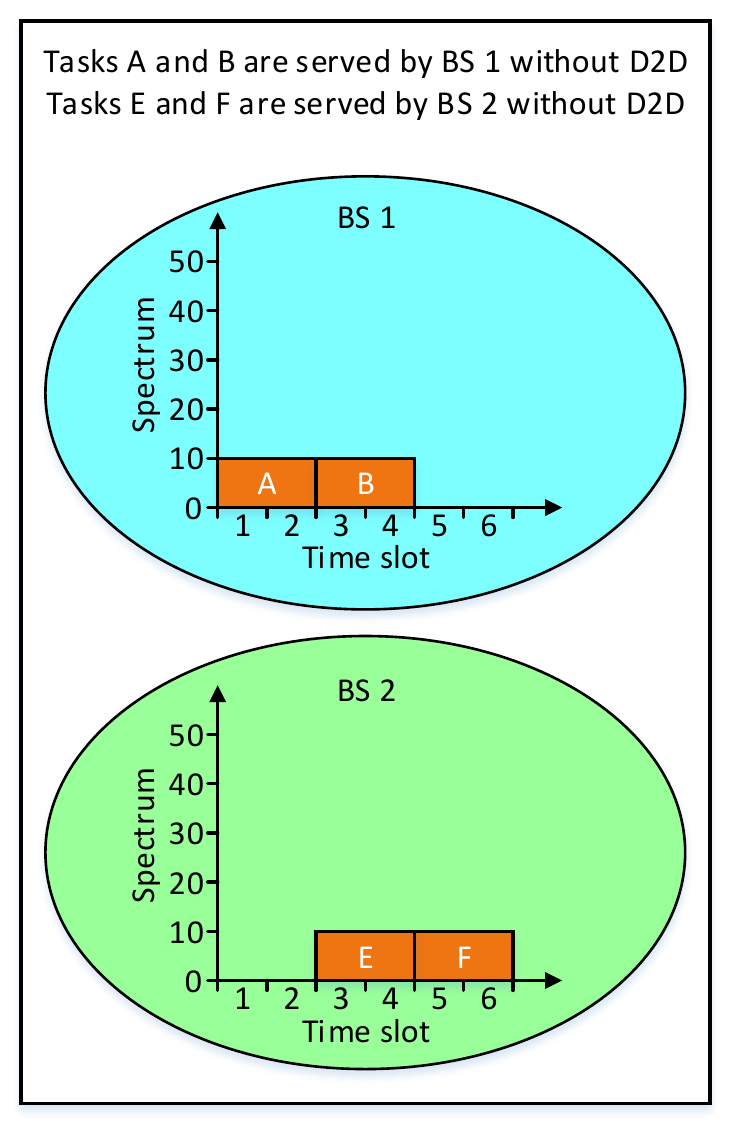}}
  \subfigure[Step III]{
    \label{fig:heuristic_Part3} 
    \includegraphics[width=0.3\linewidth]{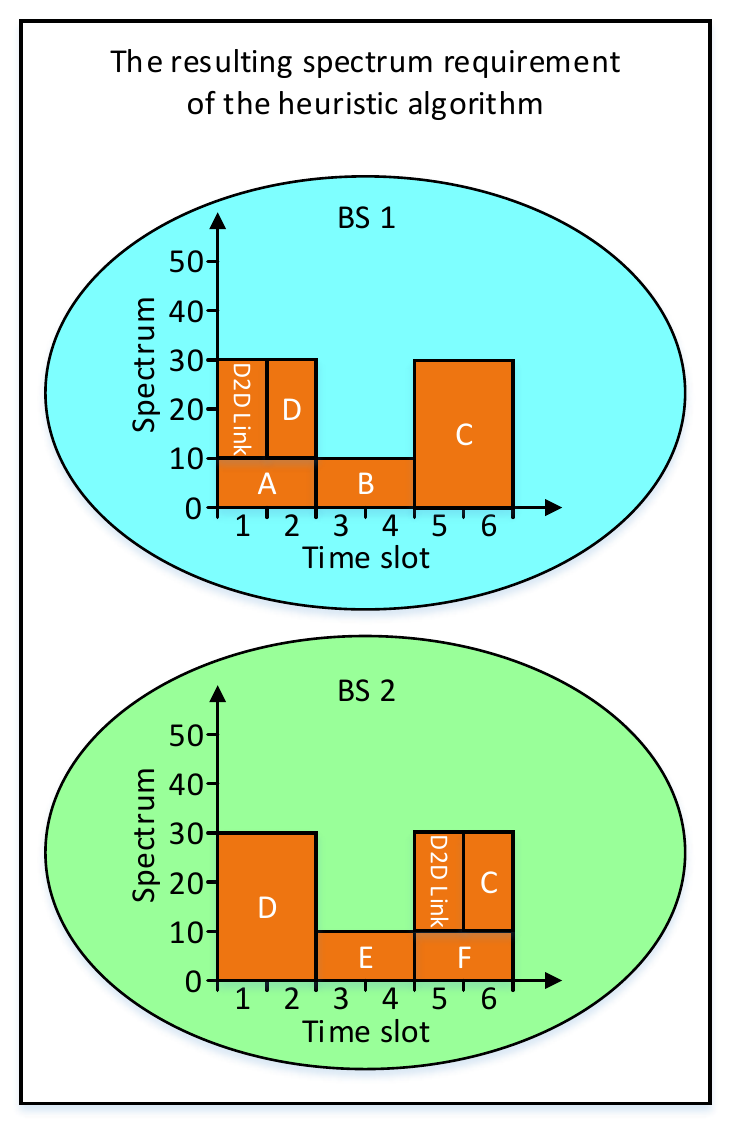}}
  \caption{An example for the heuristic algorithm. BS 1 has three tasks:
  task $A$ is generated at slot 1 and the deadline is slot 2 and the volume is 20;
  task $B$ is generated at slot 3 and the deadline is slot 4 and the volume is 20;
  task $C$ is generated at slot 5 and the deadline is slot 6 and the volume is 80.
  BS 2 has three tasks:
  task $D$ is generated at slot 1 and the deadline is slot 2 and the volume is 80;
  task $E$ is generated at slot 3 and the deadline is slot 4 and the volume is 20;
  task $F$ is generated at slot 5 and the deadline is slot 6 and the volume is 20.
  Suppose that all links have unit link rate, i.e., $R_{u,v}=1, \forall (u,v) \in \mathcal{E}$.
  }
  \label{fig:heuristic} 
\end{figure*}

We illustrate an example in Fig.~\ref{fig:heuristic} for our proposed heuristic algorithm in Sec.~\ref{sec:heuristic}.
We further analyze this example step-by-step.

  \emph{Step I.} In step I, both BSs serve their own traffic demands locally without D2D,
  whose optimal solution is shown in (a). Namely, the (peak) spectrum requirement for BS 1 is $F_1=40$ and
  the (peak) spectrum requirement for BS 2 is also $F_2=40$.

  \emph{Step II.} In step II, we take $\lambda = 0.5$. Then we can see that
  BS 1's spectrum requirement at slots 5 and 6 is larger than $\lambda F_1$ when serving task C. Thus, $\mathcal{J}_1^{\textsf{D2D}}(\lambda)=\{C\}$
  and $\mathcal{J}_1^{\textsf{ND}}(\lambda)=\{A,B\}$. Similarly, for BS 2, we have  $\mathcal{J}_1^{\textsf{D2D}}(\lambda)=\{D\}$
  and $\mathcal{J}_1^{\textsf{ND}}(\lambda)=\{E,F\}$. Then tasks A and B in
  $\mathcal{J}_1^{\textsf{ND}}(\lambda)$ are locally served by BS 1 without D2D and tasks E and F in
  $\mathcal{J}_2^{\textsf{ND}}(\lambda)$ are locally served by BS 2 without D2D, according to the optimal solution in Step I, as shown in (b).

  \emph{Step III.} In step III, task C in $\mathcal{J}_1^{\textsf{D2D}}(\lambda)$ and task D in $\mathcal{J}_2^{\textsf{D2D}}(\lambda)$  participate in D2D load balancing and are jointly served by both BS 1 and BS 2. We solve the new LP \eqref{equ:d2d-lp-heuristic}
  for tasks C and D with the already allocated spectrum in Step II for tasks A, B, E and F into consideration. The
  resulting spectrum requirement for both BSs at each slot is shown in (c).

  As we can see, as compared to solving the original problem $\textsf{Min-Spectrum-D2D}$ with 6 tasks (A-F),
  our heuristic algorithm only needs to solve the new LP \eqref{equ:d2d-lp-heuristic} with 2 tasks (C and D) with D2D load balancing,
  which reduces the computational complexity.

\fi

\section{Proof of Theorem \ref{thm:performance-heuristic}} \label{app:proof-performance-heuristic}
It is obvious that $\rho^{\textsf{Heuristic}} \le \rho$. To show
$\rho^{\textsf{Heuristic}} \ge (1 - \lambda) \rho$,
we need to show that
\be
F^{\textsf{Heuristic}}  \le  (1-\lambda)F^{\textsf{D2D}} + \lambda F^{\textsf{ND}}.
\ee

In the following, we construct a feasible solution to \eqref{equ:d2d-lp-heuristic} whose total spectrum requirement
is at most $(1-\lambda)F^{\textsf{D2D}} + \lambda F^{\textsf{ND}}$. Then since $F^{\textsf{Heuristic}} $
is the optimal  value of  \eqref{equ:d2d-lp-heuristic}, we clearly have that
$F^{\textsf{Heuristic}}  \le  (1-\lambda)F^{\textsf{D2D}} + \lambda F^{\textsf{ND}}.$

All jobs in $\mathcal{J}_b^{\textsf{ND}}(\lambda)$ are served locally according  to the results in Step I, i.e., $\{x^{j}_{u_j,b}(t)\}$.
Each job $j$ in the demand set $\mathcal{J}_b^{\textsf{D2D}}(\lambda)$ is served as follows.
For all slots not in $T_b(\lambda)$, we serve it according to the results in Step I, i.e., $\{x^{j}_{u_j,b}(t)\}$.
Thus, the resulting total spectrum for any slot $t \notin T_b(\lambda)$ is $\gamma_b(t) \le \lambda F_b$,
where $F_b$ is the optimal value of \textsf{Min-Spectrum-ND}$_b$.
At any slot $t \in T_b(\lambda)$, if job $j$ is delivered at the volume of $v$ at slot $t$ when
solving \textsf{Min-Spectrum-ND}$_b$ in Step I of our heuristic algorithm, we serve
job $j$ at the volume of $\lambda v$ locally without D2D, i.e., directly sending $\lambda v$ amount
to BSs. Thus, any job $j \in \mathcal{J}_b^{\textsf{D2D}}(\lambda)$ will be served \emph{at least} at the volume of $\lambda r_j$
locally. And the resulting total used spectrum from users to BS $b$ at slot $t \in T_b(\lambda)$ is at most $\lambda F_b$.
In summary, the resulting total spectrum for any slot $t \in [T])$ is at most $\sum_{b \in \mathcal{B}} \lambda F_b = \lambda F^{\textsf{ND}}$.

After that, every job $j \in \mathcal{J}_b^{\textsf{D2D}}(\lambda)$ has a remaining volume of at most $(1-\lambda) r_j$, i.e.,
scaling with a factor $(1-\lambda)$.
We then serve all those jobs in $\mathcal{J}_b^{\textsf{D2D}}(\lambda)$ with the remaining volume with
D2D by solving the problem $\textsf{Min-Spectrum-D2D}$. The resulting total spectrum for any slot $t \in [T]$ is at most
$(1-\lambda) F^{\textsf{D2D}}$.

Thus, the total spectrum of this constructed solution is at most  $(1-\lambda)F^{\textsf{D2D}} + \lambda F^{\textsf{ND}}$,
which completes the proof.

\ifx \ISTR \undefined
\else

\section{Proof of Theorem~\ref{the:overhead-upper-bound-heuristic}} \label{app:proof-of-overhead-upper-bound-heuristic}
According to our heuristic algorithm, only those demands in $\mathcal{J}^{\textsf{D2D}}$
can participate in D2D communication.
Since each bit of traffic demand $j$ can at most travel $e_j-s_j$ times over D2D links before it
reaches a BS. Thus, each traffic demand $j$ can at most incur D2D traffic $r_j(e_j-s_j) \le r_j (d_{\max}-1)$.
The total D2D traffic is thus upper bounded by
\be
V^{\textsf{D2D}} \le (d_{\max}-1) \sum_{j \in \mathcal{J}^{\textsf{D2D}}} r_j.
\ee
And thus the overhead ratio is upper bounded by
\bee
\eta & = \frac{V^{\textsf{D2D}}}{V^{\textsf{D2D}} + V^{\textsf{BS}}} \nnb \\
& \le \frac{(d_{\max}-1) \sum_{j \in \mathcal{J}^{\textsf{D2D}}} r_j}{(d_{\max}-1) \sum_{j \in \mathcal{J}^{\textsf{D2D}}} r_j +
\sum_{j \in \mathcal{J}} r_j} \nnb \\
& \le \frac{d_{\max}-1}{d_{\max}}.
\eee

\fi

\bibliographystyle{IEEEtran}
\bibliography{ref}

\begin{IEEEbiography}[{\includegraphics[width=1in,height=1.25in,clip,keepaspectratio]{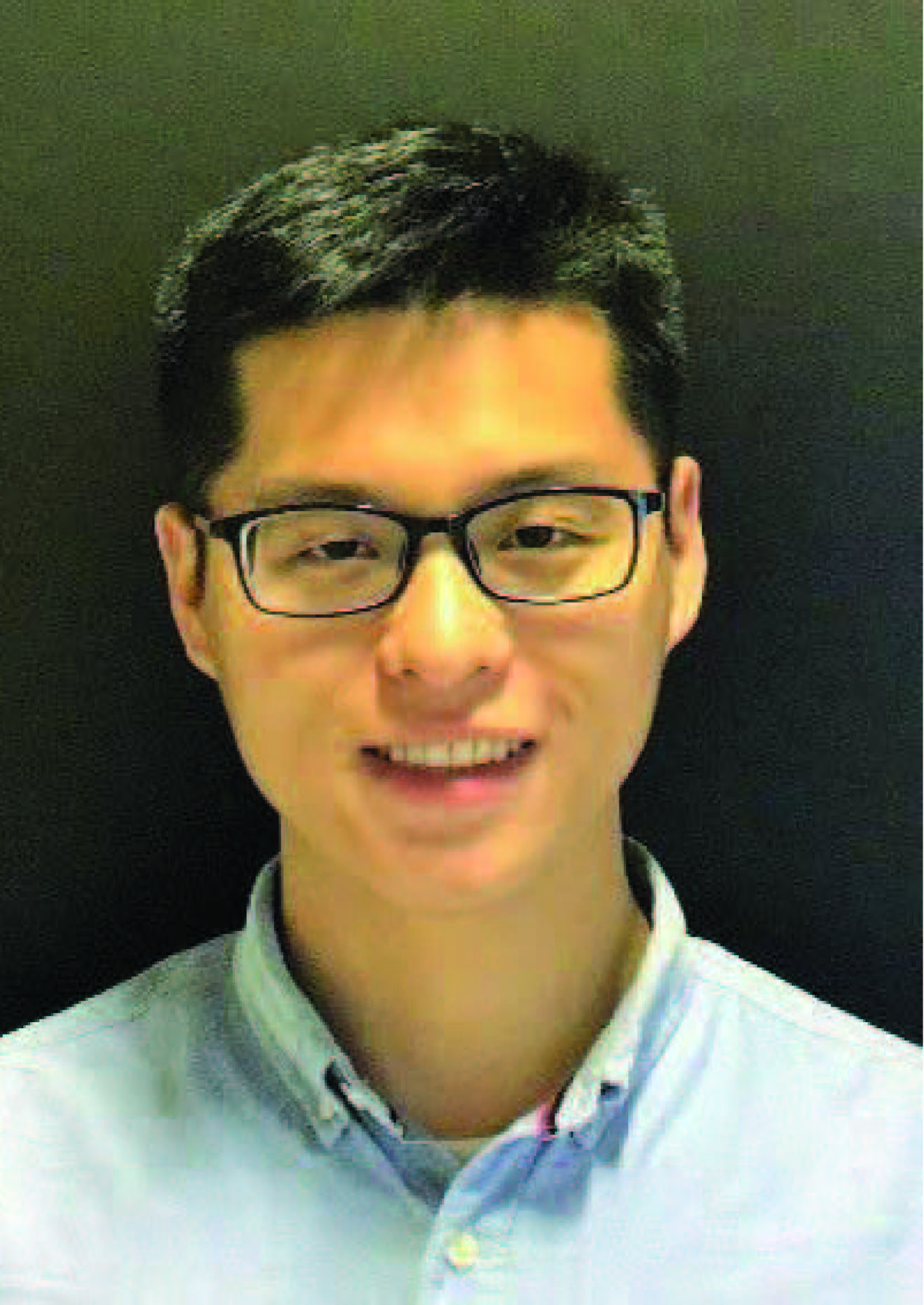}}]{Lei Deng}(S'14-M'18)
 received the B.Eng. degree from the Department of Electronic Engineering, Shanghai Jiao Tong University, Shanghai, China, in 2012, and the Ph.D. degree from the Department of Information Engineering, The Chinese University of Hong Kong, Hong Kong, in 2017. In 2015, he was a Visiting Scholar with the School of Electrical and Computer Engineering, Purdue University, West Lafayette, IN, USA. He is now an assistant professor in School of Electrical Engineering \& Intelligentization, Dongguan University of Technology. His research interests are timely network communications, intelligent transportation system, and spectral-energy efficiency in wireless networks.
\end{IEEEbiography}

\begin{IEEEbiography}[{\includegraphics[width=1in,height=1.25in,clip,keepaspectratio]{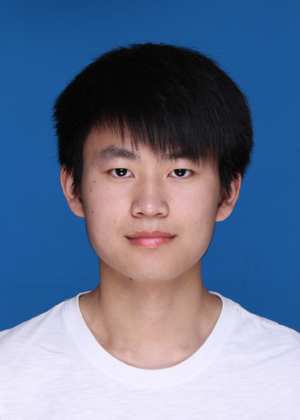}}]{Yinghui He}
 received the B.S.E. degree in information engineering from Zhejiang University, Hangzhou, China, in 2018. He is currently pursuing the master's degree with the College of Information Science and Electronic Engineering, Zhejiang University. His research interests mainly include mobile edge computing and device-to-device communications.
\end{IEEEbiography}

\begin{IEEEbiography}[{\includegraphics[width=1in,height=1.25in,clip,keepaspectratio]{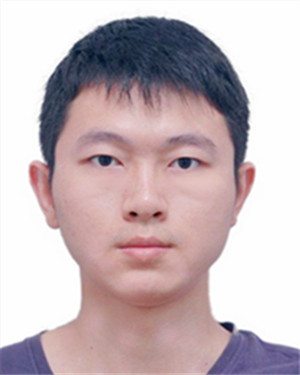}}]{Ying Zhang}
received his B.Eng. degree from the Department of Electronic Engineering, Shanghai Jiao Tong University, Shanghai, China, in 2013.
He received his Ph.D. degree from the Department of Information Engineering, The Chinese University of Hong Kong, Hong Kong, in 2017.
His research interests include energy system operation and optimization, machine learning, statistical arbitrage and algorithmic trading.
\end{IEEEbiography}

\begin{IEEEbiography}[{\includegraphics[width=1in,height=1.25in,clip,keepaspectratio]{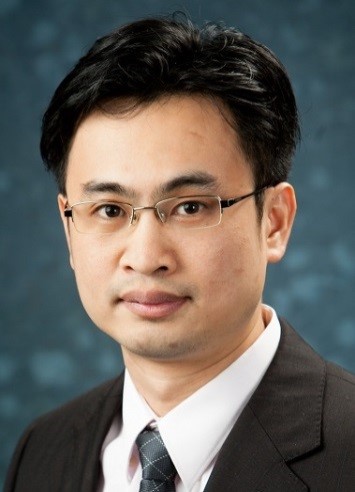}}]{Minghua Chen} (S'04-M'06-SM'13) received his B.Eng. and M.S. degrees from the Department of Electronic Engineering at Tsinghua University in 1999 and 2001, respectively. He received his Ph.D. degree from the Department of Electrical Engineering and Computer Sciences at University of California at Berkeley in 2006. He spent one year visiting Microsoft Research Redmond as a Postdoc Researcher. He joined the Department of Information Engineering, the Chinese University of Hong Kong, in 2007, where he is now an Associate Professor. He is also currently an Adjunct Associate Professor in Tsinghua University, Institute of Interdisciplinary Information Sciences. He received the Eli Jury award from UC Berkeley in 2007 (presented to a graduate student or recent alumnus for outstanding achievement in the area of Systems, Communications, Control, or Signal Processing) and The Chinese University of Hong Kong Young Researcher Award in 2013. He also received several best paper awards, including the IEEE ICME Best Paper Award in 2009, the IEEE Transactions on Multimedia Prize Paper Award in 2009, and the ACM Multimedia Best Paper Award in 2012. He is currently the Steering Committee Chair of ACM e-Energy. He serves as TPC Co-Chair of ACM e-Energy 2016 and General Chair of ACM e-Energy 2017. He also serves as Associate Editor of IEEE/ACM Transactions on Networking in 2014-2018. He receives the ACM Recognition of Service Award in 2017 for service contribution to the community. His recent research interests include energy systems (e.g., smart power grids and energy-efficient data centers), intelligent transportation systems, distributed optimization, multimedia networking, wireless networking, delay-constrained networking, and characterizing the benefit of data-driven prediction in algorithm/system design.
\end{IEEEbiography}

\begin{IEEEbiography}[{\includegraphics[width=1in,height=1.25in,clip,keepaspectratio]{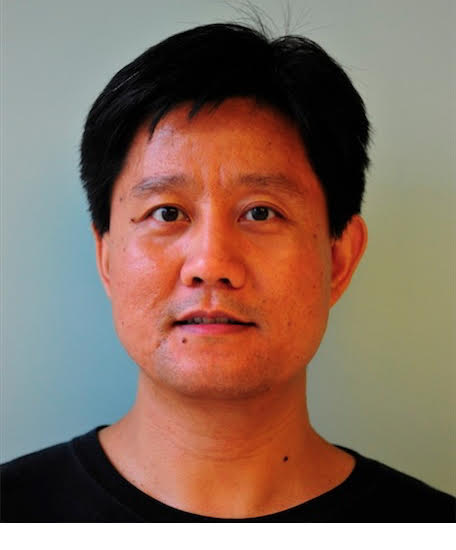}}]{Zongpeng Li}
received his BSc in Computer Science from Tsinghua University in 1999, and his PhD from University of Toronto in 2005. He was affiliated with University of Calgary, and is now a Professor at the School of Computer Science, Wuhan University. His research interests include computer networks, cloud computing, and IoT.
\end{IEEEbiography}

\begin{IEEEbiography}[{\includegraphics[width=1in,height=1.25in,clip,keepaspectratio]{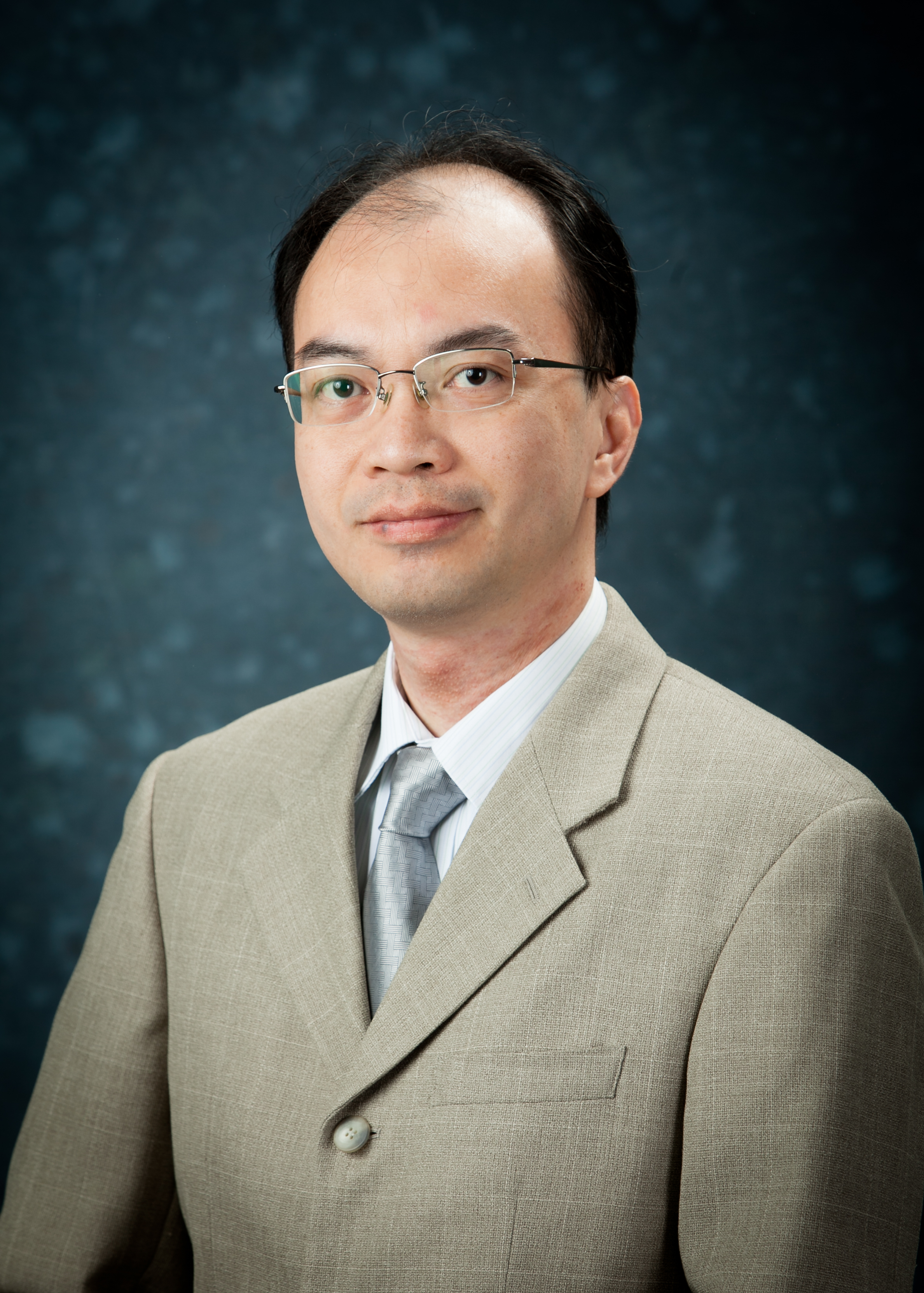}}]{Jack Y. B. Lee} (M'95-SM'03)
received his B.Eng. and Ph.D. degrees in Information Engineering from the Chinese University of Hong Kong, Shatin, Hong Kong, in 1993 and 1997, respectively. He is currently an Associate Professor with the Department of Information of the Chinese University of Hong Kong. His research group focuses on research in multimedia communications systems, mobile communications, protocols, and applications. He specializes in tackling research challenges arising from real-world systems. He works closely with the industry to uncover new research challenges and opportunities for new services and applications. Several of the systems research from his lab have been adopted and deployed by the industry.
\end{IEEEbiography}

\begin{IEEEbiography}[{\includegraphics[width=1in,height=1.25in,clip,keepaspectratio]{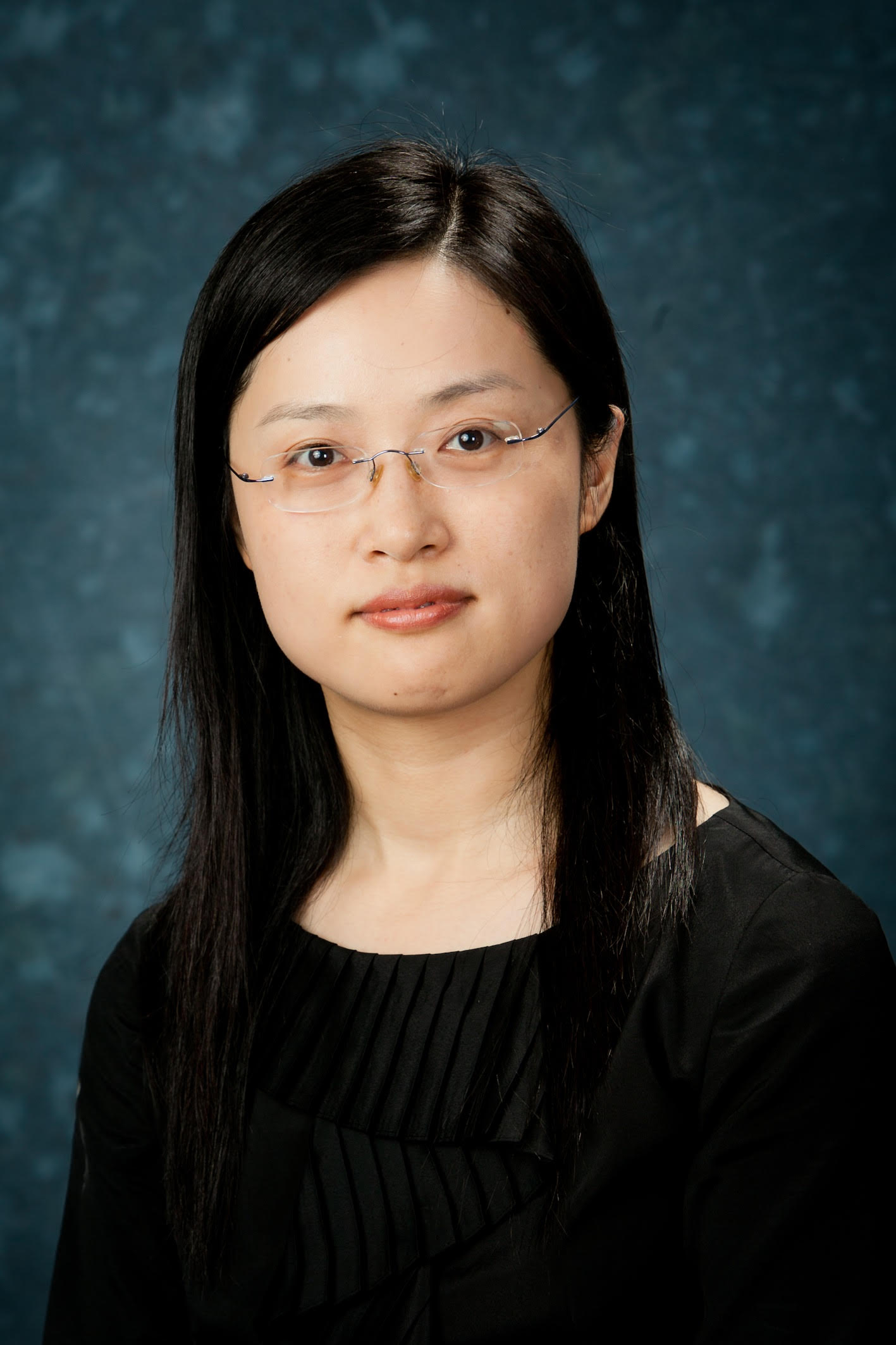}}]{Ying-Jun Angela Zhang} (S'00-M'05-SM'10)
received her PhD degree in Electrical and Electronic Engineering from the Hong Kong University of Science and Technology, Hong Kong in 2004. Since 2005, she has been with Department of Information Engineering, The Chinese University of Hong Kong, where she is currently an Associate Professor. Her research interests include mainly wireless communications systems and smart power systems, in particular optimization techniques for such systems.
She serves as the Chair of the Executive Editor Committee of the IEEE Transactions on Wireless Communications. Previously, she served many years as an Associate Editor of the IEEE Transactions on Wireless Communications, IEEE Transactions on Communications, Security and Communications Networks (Wiley), and a Feature Topic in the IEEE Communications Magazine. She has served  on the organizing committee of major IEEE conferences including ICC, GLOBECOM, SmartgridComm, VTC, CCNC, ICCC, MASS, etc.. She is now the Chair of IEEE ComSoc Emerging Technical Committee on Smart Grid. She was a Co-Chair of the IEEE ComSoc Multimedia Communications Technical Committee and the IEEE Communication Society GOLD Coordinator. She was the co-recipient of the 2014 IEEE ComSoc APB Outstanding Paper Award, the 2013 IEEE SmartgridComm Best Paper Award, and the 2011 IEEE Marconi Prize Paper Award on Wireless Communications. She was the recipient of the Young Researcher Award from the Chinese University of Hong Kong in 2011. As the only winner from engineering science, she has won the Hong Kong Young Scientist Award 2006, conferred by the Hong Kong Institution of Science. Dr. Zhang is a Fellow of IET and a Distinguished Lecturer of IEEE ComSoc.
\end{IEEEbiography}

 \begin{IEEEbiography}[{\includegraphics[width=1in,height=1.25in,clip,keepaspectratio]{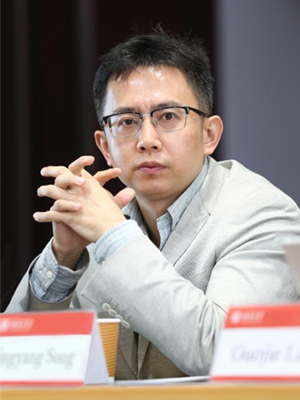}}]{Lingyang Song} (S'03-M'06-SM'12)
received his PhD from the University of York, UK, in 2007, where he received the K. M. Stott Prize for
excellent research. He worked as a research fellow at the University
of Oslo, Norway until rejoining Philips Research UK in March 2008. In
May 2009, he joined the School of Electronics Engineering and Computer
Science, Peking University, and is now a Boya Distinguished Professor.
His main research interests include wireless communication and
networks, signal processing, and machine learning. He is the recipient
of IEEE Leonard G. Abraham Prize in 2016 and IEEE Asia Pacific (AP)
Young Researcher Award in 2012. He is a senior member of IEEE, and an
IEEE distinguished lecturer since 2015.
\end{IEEEbiography}

\end{document}